# Omni-tomography / Multi-tomography
## – Integrating Multiple Modalities for Simultaneous Imaging

Ge Wang[1], Jie Zhang[2], Hao Gao[3], Victor Weir[4], Hengyong Yu[5], Wenxiang Cong[1], Xiaochen Xu[6], Haiou Shen[1], James Bennett[1], Yue Wang[7], Michael Vannier[8]

[1] Biomedical Imaging Division, VT-WFU School of Biomedical Engineering and Sciences, Blacksburg, Virginia Tech, Blacksburg, Virginia, USA

[2] Department of Radiology, University of Minnesota, Minneapolis, Minnesota, USA

[3] Department of Mathematics, University of California, Los Angeles, California, USA

[4] Department of Medical Physics and Radiation Safety, Baylor University Medical Center, Dallas, Texas, USA

[5] Biomedical Imaging Division, VT-WFU School of Biomedical Engineering and Sciences, Wake Forest University Health Sciences, Winston-Salem, North Carolina, USA

[6] Medical Business Unit, Texas Instruments Inc., Dallas, Texas, USA

[7] Department of Electrical and Computer Engineering, Virginia Tech, Blacksburg, Virginia, USA

[8] Department of Radiology, University of Chicago, Chicago, Illinois, USA

ge-wang@ieee.org, jzctjmn@gmail.com, haog@math.ucla.edu, vjweir@gmail.com, hyu@wfubmc.edu, congw@vt.edu, xiaochenxu@ti.com, hhshen@vt.edu, jrbennet@gmail.com, yuewang@vt.edu, mwvannier@gmail.com

June 10, 2011

**Abstract:** Current tomographic imaging systems need major improvements, especially when multi-dimensional, multi-scale, multi-temporal and multi-parametric phenomena are under investigation. Both preclinical and clinical imaging now depend on *in vivo* tomography, often requiring separate evaluations by different imaging modalities to define morphologic details, delineate interval changes due to disease or interventions, and study physiological functions that have interconnected aspects. Over the past decade, fusion of multimodality images has emerged with two different approaches: post-hoc image registration and combined acquisition on PET-CT, PET-MRI and other hybrid scanners. There are intrinsic limitations for both the post-hoc image analysis and dual/triple modality approaches defined by registration errors and physical constraints in the acquisition chain. We envision that tomography will evolve beyond current modality fusion and towards grand fusion, a large scale fusion of all or many imaging modalities, which may be referred to as omni-tomography or multi-tomography. Unlike modality fusion, grand fusion is here proposed for truly simultaneous but often localized reconstruction in terms of all or many relevant imaging mechanisms such as CT, MRI, PET, SPECT, US, optical, and possibly more. In this paper, the technical basis for omni-tomography is introduced and illustrated with a top-level design of a next generation scanner, interior tomographic reconstructions of representative modalities, and anticipated applications of omni-tomography.

## I. Background

### A. Introduction

The physiome concept was first presented to the International Union of Physiological Sciences (IUPS) in 1993, and later designated as a strategic area by IUPS in 2001[1-3]. Physiome describes physiological processes and their interactions from the genome scale to complex organism in a systematic fashion. The IUPS Physiome Project supports a worldwide repository of models and datasets, and represents an integral component of systems biology and modern medicine. Medical imaging has a huge potential to support the physiome project with a yet-to-be-much-improved capability of obtaining functional information at multi-levels from organs to physiological units.

In the medical imaging field, efforts are being made to link molecular assays with diagnostic imaging [4-5]; however, success to date has been rather limited. One reason is that current medical imaging scanners do not individually offer a wide enough spectrum of information. For example, current x-ray CT scanners produce gray-scale images of physiological or pathological process, On the other hand, large amounts of information from genetic and epigenetic profiling is becoming available. This imbalance between phenotype information (e.g., CT images) and genome-level information (e.g., RNA data) demands more features from the *in vivo* imaging side. Indeed, the medical imaging field is rapidly trending in this direction. Turning again to x-ray CT, we see transition from gray-scale to true-color images with development of energy-sensitive, photon-counting detector technology. Another area of advancement is x-ray phase-contrast and dark-field imaging. Overall, imaging modalities and contrast agents are constantly being improved to generate increasingly more information about structural, functional, cellular and molecular characteristics.

The holy grail of biomedical imaging would be an integrated system capable of producing simultaneous, dynamic observations *in vivo* of highly complex biological phenomena. The multimodal fusion approach has made significant strides towards meeting this challenge, as demonstrated by the popularity of PET-CT and other hybrid systems [6-10]. We envision that the next stage will be integrating ever more modalities into a single scanner for truly simultaneous and information-rich data acquisition, including CT, MRI, PET, SPECT, US imaging, optical imaging, and more. In the following subsections we will examine the current status of modality fusion and future directions towards omni-tomography.

### B. Individual Modalities

Projection x-ray imaging revolutionized medical diagnostics and inspired the development of other tomographic imaging modalities. There is now an impressive array of scanners for computed tomography (CT), magnetic resonance imaging (MRI), positron emission tomography (PET) and single photon emission computed tomography (SPECT), ultrasonic imaging (US) imaging, optical imaging, and more. These modalities utilize different physical principles and reveal information from distinct but complementary perspectives. A recent book chapter on the state-of-the-art in medical imaging, from several of the coauthors in this article, provides further details and references [11]. Next, we will present a brief synopsis of popular imaging modalities along with Table I that summarizes key features of these modalities.

X-ray CT (henceforth CT) has been rather popular over the past two decades: about 100 million scans are performed annually in the United States alone. With inherently higher spatial resolution (about 0.3mm) and faster imaging speeds (about 100 ms) [12], CT can examine tissues that differ in density by less than 1%, along with physiological and pathological dynamics. A single circular CT scan can cover many important biomedical targets with its large longitudinal range (up to 16 cm). Moreover, a spiral cone-beam CT scan further extends this coverage to solve a long object problem (e.g., whole body angiography). The primary shortcomings of current CT scanners are patient radiation dose and limited contrast resolution, both of which can be significantly improved using contrast agents.

When compared with CT, MRI is distinguished by its superior contrast resolution. Unlike CT imaging, MRI does not involve ionizing radiation and has multiple contrast mechanisms: proton density weighted, T1 or T2 weighted, contrast-enhanced, motion sensitive, elastic, temporal, and chemical shift imaging [13-19]. However,

MRI is generally considered significantly slower and more expensive than CT. In addition, scanning patients with certain implantable devices is contraindicated unless the device is MRI compatible [20].

PET has the advantages of high sensitivity and specificity with moderate spatial resolution (one to several millimeters) [21] [22] [23]. It is heavily used in clinical oncology, brain mapping, neurologic and cardiac functional imaging. While CT and MRI isolate anatomic and functional changes, PET (and SPECT) are capable of detecting molecular activities, even prior to phenotype expressions [24-25]. PET uses injected radiochemical probes (positron emitters), with differential uptake rates modulated by tissue vasculature and metabolism, and takes longer scan time (~30 minutes) than CT and MRI. Quantitative information can be extracted from PET. The major limitation to its widespread use is the high cost of producing its short-lived radionuclides and radiopharmaceutical probes.

SPECT, along with PET, depends on radioactive probes and resultant gamma rays [26-27]. In contrast to PET, SPECT probes emit gamma photons that are directly measurable, whereas PET probes emit positrons that annihilate with electrons to form a pair of measurable collinear gamma photons. Generally, SPECT has a lower spatial resolution (about 10mm) and a longer scan time (several hours) than PET, but it is significantly less expensive than PET, and is important for cardiac, brain, tumor, infection, thyroid or bone imaging.

Ultrasound (US) imaging stands alongside CT, MRI, PET and SPECT, and has profoundly impacted the medical practice due to its portability, safety, cost effectiveness, and real-time performance [28]. However, medical US is limited by the high attenuation and complexity of acoustic interaction with bone and air. Thus, it is primarily useful for examining the outer surface of soft structures and organs. US images can be enhanced with contrast agents such as micro-bubbles. On the other hand, focused US can serve therapeutic purposes.

Optical imaging depends on optical parameters and/or light probes [29]. Diffuse optical tomography (DOT) reconstructs distributions of tissue absorption and scattering coefficients, and has applications in breast imaging *etc*. Enabled by fluorescence probes, optical fluorescence imaging can sense biological processes *in vivo* at the cellular and molecular levels. Of particular interest, fluorescence tomography employs fluorescence signals, induced by laser light or x-rays, to determine a volumetric distribution of fluorescent probes. As a prerequisite, anatomical structures and optical properties of the tissue need to be estimated.

*Table I* *Comparison of medical imaging modalities CT, MRI, PET, SPECT, US imaging and optical imaging.*

| Imaging Modality | Spatial/Temporal Resolution | Imaging Mechanism | Strengths | Drawbacks |
|---|---|---|---|---|
| CT | 10nm-1mm/10ms | X-ray attenuation | High spatial and temporal resolution, high bony contrast | Ionizing radiation, low soft-tissue contrast, less functional and biochemical information content |
| MRI | 100µm-5mm/1s | Magnetization of atoms | High soft-tissue contrast, rich functional, biochemical, and metabolism information content | Lower spatial and temporal resolution, body device and implants incompatibility |
| PET | 1mm/100s | Positron-labeled ligand binding | High sensitivity and specificity, rich functional, biochemical, and molecular information content | Lower spatial and temporal resolution, partial-volume effect, high noise |
| SPECT | 1mm-10mm/100s | Gamma-emitting radioligand binding | High sensitivity and specificity, rich functional, biochemical, and molecular information content | Low spatial and temporal resolution, partial-volume effect, high noise |
| US Imaging | 100µm-1mm/10ms | Acoustic-tissue interaction | Real-time imaging, compact and portable | Difficult through air and bone, limited field of view, less functional and biochemical information content |
| Optical Imaging | 1mm-10mm/1s | Target-seeking light-emitting probing | Compact, rich molecular information content | Limited penetration, poor spatial and temporal resolution |

## C. Modality Fusion

Since the advent of diagnostic imaging, there has been a powerful push to combine imaging modalities in a single global coordinate system. The results are widely used multimodality commercial systems: PET-CT, SPECT-CT PET-MRI, and others. These systems represent the cutting-edge of imaging technology, are generally well received and have greatly facilitated important biomedical imaging tasks [7-9, 30-38].

The area of modality fusion began with PET-CT which revolutionized medical diagnosis. A PET-CT scanner sequentially acquires patient PET and CT images in an integrated gantry. As a result, functional information (e.g. metabolic or biochemical processes) obtained from PET can be accurately co-registered with anatomic information from CT. As such, PET-CT has forever changed the fields of oncology, surgical planning, radiation therapy, and others so much that most imaging centers have upgraded from conventional PET to PET-CT.

Examples of contemporary modality fusion are not rare. Mediso developed the first human PET, SPECT and CT system AnyScan (http://www.mediso.de/anyscan-sc.html). As shown in **Figure I.C.1(a)**, it provides sequential anatomical and functional imaging within a single framework, accommodating all images with motion artifact suppression and attenuation background correction. Another company, Carestream, has put seven preclinical imaging modalities in two instruments (http://www.cmi-marketing.com/7modalities), covering PET, SPECT, CT, fluorescence, luminescence, radioisotopic and radiographic imaging. As shown in **Figure I.C.1(b)**, the Carestream Albira system is designed for sequential micro-PET, micro-SPECT and micro-CT data acquisitions and global reconstructions, similar to the Mediso AnyScan system.

PET-MRI is the most recent frontier in modality fusion [30-31, 39]. PET/MRI delivers molecular information (e.g., cell surface reactors, enzymes and gene expression) from PET, along with anatomical and functional data from MRI. Hence, PET/MRI is capable of assessing flow, diffusion, perfusion and cardiac motion in a single examination, and evaluating myocardial viability (PET) and metabolism (MRI) for diagnosis of cardiac pathology, such as coronary artery disease.

## D. Grand Fusion

To go beyond existing modality fusions, it might be imagined that additional modalities should be added for simultaneous characterization of additional biomedical properties [34, 40-41]; however, this mission may appear impractical due to space conflict with the physical size requirements of current scanners and other physical constraints. We could longitudinally assemble additional scanners, but this sequential arrangement would make synchronized capture impossible, especially when relatively slow modalities are involved (e.g. PET and SPECT). The arguments for and limitations of the classic modality fusion approach are demonstrated in the latest development of the Advanced Multimodality Image Guided Operating (AMIGO) Suite project, shown in **Figure I.D.1**.

To overcome the aforementioned challenge for grand fusion, let us elevate interior tomography from a specific imaging mode to a general guiding principle for the biomedical imaging field. Over the past years, interior tomography has been studied for theoretically exact CT image reconstruction over an internal region of interest (ROI) from data associated only with lines through the ROI, which means that a relatively narrow imaging chain can be made [42-44]. This approach has been extended for interior SPECT [45], and also achieved some success for interior MRI [46]. We propose to compress each imaging modality into a slim imaging chain for ROI-targeted reconstruction, instead of traditional global reconstruction, and use a single gantry system for truly concurrent signal acquisition and composite interior reconstruction in a unified framework, regularized by prior knowledge from sparsity to atlases. We call this paradigm shift omni-tomography or multi-tomography.

We believe that omni-tomography will open up exciting new opportunities in research, development and applications. Specifically, it may offer temporally-correlated quantitative physiological, pathological, pharmacokinetic and pharmacodynamic information. Examples include: metabolic processes, enzyme activities, protein bindings, kinetic interplays, tumor & stem cell tracking, longitudinal studies in cardiology, oncology, neuroscience, tissue engineering, and so on. The fundamental implications of omni-tomography will be further discussed in the last section of this paper.

### E. Outline

In section II, we describe the first of its kind top-level design of an omni-tomographic imaging system. We considered a number of variants and realized that each architectural style has advantages and disadvantages. Our collective agreement is on an "O" design which represents an optimal balance between imaging performance and cost-effectiveness. In this feasibility study, we focus on primary features and system integration. Needless to say, there are numerous details yet to be specified.

In section III, we report preliminary results on interior reconstructions using representative imaging modalities including CT, MRI, SPECT, and x-ray fluorescence tomography. While we have peer-reviewed papers on interior CT and interior SPECT, interior MRI based on an inhomogeneous magnetic background field and interior x-ray fluorescence tomography based on novel nanoparticles are regarded as two novel contributions by themselves.

Finally, in section IV, we discuss the utilities and necessities of omni-tomography from multiple perspectives. Specifically, omni-tomography, or grand fusion, is hereby proposed as the next stage beyond modality fusion. Omni-tomography offers great synergy *in vivo* for diagnosis, intervention, and drug development, can be made versatile and cost-effective, and as such become an unprecedented imaging platform for development of systems biology and modern medicine. Given this magnitude of capabilities and complexities, there are endless opportunities ahead.

## II. System Architecture

### A. General Description

Given the physical requirements of typical scanners, the primary challenge is to integrate all or many of them into a single gantry. Thanks to the interior tomography principle, we have substantial flexibility to integrate various imaging modalities when they only target a relatively small ROI. We have systematically analyzed a number of architectures and realized that each style has advantages and disadvantages. Here we focus on a ring-shaped design, which only serves as an initial example of omni-tomography instrumentation.

**Figure II.A.1** illustrates the proposed system architecture for omni-tomography. For the first time, it integrates all the major imaging modalities into a conventional gantry space for truly simultaneous acquisition of CT, MRI, PET, SPECT, Ultrasound and optical imaging. This top-level design is consistent with the governing physical laws, and demonstrates the feasibility of omni-tomography. Given the page limitation, we will emphasize the primary features instead of the secondary technical details.

All the major medical tomographic modalities are incorporated into three concentric rings: an inner ring as a permanent magnet; a middle ring containing the x-ray tube and detector array, and a pair of SPECT detectors; and an outer ring for PET crystals and electronics (inner and outer rings are static). The middle ring is designed to rotate and acquire data for both interior CT and interior SPECT. This rotating ring is embedded in a slip-ring (similar to a large diameter ball bearing) which supports the rotating ring and facilitates power/signal transmission. The rotating ring, the slip-ring, and the PET ring all go through the magnetic poles. The yoke for N and S poles of the magnet are configured similarly to a "C-arm". The system is designed for human or animal subjects, and can accommodate a standard patient size (approximately 170 cm in height, 70 kg, with a chest size of 22 cm in AP direction and 35 cm in lateral direction). The key features specific to each imaging modality follows next.

### B. Specific Remarks

**MRI –** The MRI component is similar to that of a commercial open MRI. All the available techniques for open MRI can be potentially adapted for our proposed omni-tomographic scanner. As shown in **Figure II.A.1**, the MRI subsystem consists of two permanent magnet heads at each magnetic pole. The vertical gap between the magnet poles is 50 cm and chosen based on a simulation to provide a sufficient magnetic field for a region of interest (ROI) of approximately 15 to 20 cm in the center of the gantry aperture; the ROI was also used to determine the width (40cm) and length (2 x 40cm) of the magnetic heads. This configuration leaves sufficient space for other modalities to probe the subject without being significantly blocked by the magnet. A deviation from the commercial open MRI design is that each magnet head is hollow and has a gap to let the middle ring modalities "look through" the magnet. Hence, the CT tube and detector, as well as the SPECT cameras, perform full-scans to the extent defined by the gap through the magnet, and cone-beam scans when the magnet is not in the radiation paths. Our detailed design for the magnet permits a 2 cm clearance between the two magnet parts of each pole.

We used Vizimag (http://www.vizimag.com/) software to simulate the field strength of the magnetic field and realize a locally uniform field between the poles. The desired magnetic field strength was set to 0.2 Tesla. The magnetic field may be adjusted by changing ferromagnetic materials and the dimensions of the magnetic blocks or using an alternative technology. The generated magnetic field is shown in **Figure II.B.1**. The magnetic flux varies from 0.208 to 0.211 Tesla over an ROI of 20 x 20 $cm^2$ with its origin at the iso-center of the main imaging plane of the omni-tomographic imager. The field uniformity can be further improved with technical refinements.

The gradient coils used with current open MRI scanners can be modified for integration into our omni-tomography gantry. **Figure II.A.1** shows a potential configuration with gradient coils inserted. Generally speaking, any gradient and RF coil settings for open MRI can be used in our proposed system [47-48].

MRI shielding is required for our proposed architecture, including: radio frequency interference shielding, electromagnetic interference shielding, electromagnetic pulse shielding, and so on. Highly conductive, non-woven electromagnetic shielding materials will be used for this purpose. These materials are usually made from a variety of fibers, such as carbon and nickel-coated carbon, which are flexible enough to accommodate complex contours and shapes. These techniques are well-developed or maturing, and should not pose significant difficulty in principle [49].

**CT –** The middle ring of the omni-tomography system contains the CT and SPECT subsystems. The CT subsystem has an x-ray source and opposing x-ray detector array. A typical source configuration would include an x-ray source (e.g. Varian GS-3074, 23.5 x 410 x 13.5 $cm^3$) with a heat exchanger (e.g. Varian HE 300, 23.5 x 410 x 13.5 $cm^3$) and a generator (e.g. Spellman 16010, 20.5 x 40 x 50 $cm^3$). A flat panel (e.g. Varian PaxScan4030CB, 47 x 37 x 7 $cm^3$) or photon-counting spectroscopic detector array may be used. The proposed source-to-detector distance is approximately 85 cm, which is in the range of conventional CT scanners.

Interior tomography was originally developed for CT, and has already produced impressive results [50-53]. We expect few issues with image reconstruction over an ROI from truncated projection data. Note that the interior CT scan data could be potentially used to estimate the attenuation background over the whole field of view, and enable anatomically-specific attenuation correction for SPECT and PET. In the future, a field emission x-ray source (http://www.xintek.com/products/xray/index.htm) can be used for its dynamic programming capability which is ideally suited for physiological gating and other studies. Also, a grating-based or other coded-aperture-type apparatuses may be used for phase-contrast and dark-field imaging. Time-of-flight x-ray imaging could be a distant possibility. Our design is flexible enough to incorporate these in the future.

**SPECT –** Two solid-state SPECT cameras are included in our design. The SPECT detectors are currently collimated to parallel-beam geometry and arranged orthogonally. The dual-detectors should double sensitivity and speed; solid-state CZT SPECT detectors are preferred due to their size and functionality. There are several commercially available systems (Gamma Medica, GE Triumph) that use CZT detectors for molecular imaging. A CZT detector can potentially detect both x-ray and gamma-ray photons simultaneously, which is great for future development. The 16 x 20 cm CZT SPECT detector, manufactured by Gamma Medica (Northridge, CA), was selected for our system.

A converging or pinhole collimator may be used within the SPECT subsystem. Both collimators types magnify features in a ROI. A multi-pinhole collimator is an option when better sensitivity is desired [54] [55]. A diverging collimator can image larger structures with a smaller detector. Furthermore, data compromised when the SPECT cameras are behind the magnetic heads can be fixed by attenuation correction because the fixed magnetic structures are semi-transparent to gamma rays. The rationale for positioning the SPECT and CT subsystems on the middle ring, instead of being in front of the magnetic heads, is to limit interference from the rotating parts on the magnetic field.

**PET –** The PET detector ring of 120 cm internal diameter consists of LYSO crystals, but potentially could be built out of CZT or other suitable solid-state materials. For LYSO crystal-based systems, the resulting scintillation emission can be detected by avalanche photodiode detectors (APDs). The detector units are 4 x 4 x 20 mm$^3$. There are 4 units per detector block; 471 detector blocks per ring; and 20 rings in total. The axial extent is 160 mm. Another component is a coincidence timing or time of flight (TOF) analysis circuit, which are commercially available.

The 511 KeV PET photons are sufficiently energetic to pass through the solid-state CT and SPECT detectors and reach the PET detector. The PET circuitry can be customized for ROI imaging. Interior PET reconstruction can be performed using an adapted interior tomography algorithm. Prior knowledge of the CT and SPECT detectors structures can be utilized to conduct hardware-related attenuation correction for PET. Another important consideration is the potential interference from the MRI subsystem that could degrade the performance of the PET subsystem. Positioning the PET ring furthest away from the MRI subsystem should produce maximum isolation for the PET detector and associated electronics from the influence of the magnet field.

In some existing PET-MRI systems, the PET ring is inside the main magnet ring. For example, the Siemens Biograph PET-MRI system incorporates new MRI compatible PET detectors into an MRI system. Many PET-MRI systems use PET inserts for simultaneous PET and MRI. Another system, depicted on a poster from Cambridge University (Ansorge et al.), uses two magnet coils placed end-to-end with a gap in between for the PET detectors. These designs are fundamentally constrained by the global data acquisition and image reconstruction requirements, and cannot reach the same level of hybrid imaging as omni-tomography.

**US –** US imaging is arguably the most cost-effective and widely used clinical imaging modality. The role of US imaging in omni-tomography is equally important. In our design, the US transducer can be fixed to a physiologically relevant ROI on the patient prior to the scan. MRI compatible US systems are already commercially available. In a typical system, the US transducer and cables are shielded with aluminum foil [56]. In addition to US imaging, photo-acoustic imaging is an emerging modality and seems a promising direction for future integration as the technology matures.

**Optical Imaging –** Optical imaging is a rapidly growing area and works by detecting molecular and cellular information via fluorescence and bioluminescence probes. Interestingly, x-rays and gamma rays can induce fluorescence and luminescence signals, which can be quite significant when fluorescence and luminescence nanoparticles (e.g. quantum dots and nano-phosphors) are in high concentrations. Due to the use of the interior imaging principle, there is unused space in the CT-SPECT ring where x-ray luminescence and x-ray fluorescence cameras could be potentially placed. Among the various possibilities, we would first consider interior x-ray fluorescence CT because x-rays are not much diffusive and give strong signals with well-designed nanoparticles. As a side note, additional x-ray detectors outside the primary CT beam may be utilized for scattering tomography which depicts scattering characteristics in 3D and offers information complementary to linear attenuation characteristics, especially using spectroscopic x-ray detection.

C. Cost Analysis

**System Cost Analysis –** Most clinical CT and MRI scanners roughly cost from $1.0 million to $2 million (USD) depending on the options selected. PET systems typically cost even more than CT or MRI. We believe an omni-tomography scanner that integrates all imaging modalities in one gantry will likely fall in the same cost range. Based on our initial cost analysis in **Table II**, an omni-tomography scanner may cost about $1.5 million. The estimate is based on the costs of various components and modules from different manufacturers and sources, as well as various software options and other specialty selections that are possibly additions to our omni-tomography system. In our proposed omni-tomography design, use of a slip ring technology enables data to be transmitted from various modalities while a subject is being scanned. The advantages of an omni-tomography scanner over current multimodality systems are not only valuable tomographic information within the same time window but also major space savings as well as logistic convenience.

**Potential Cost Savings –** Aside from the above estimation, we envision that omni-tomography may significantly reduce costs related to oncological imaging. Typically, patients who are diagnosed with cancer will receive multiple scans of various modalities during their course of treatment. The following data from Duke University (via the website auntminnie.com) helps illustrate this point: in 2006, the average patient with lung cancer received (within two years of diagnosis) 11 radiographs, one PET scan, six CT scans, a separate nuclear medicine test, one MRI exam, two echocardiograms, and an ultrasound. It is likely that each of these scans were performed on a separate scanner within the hospital/clinic. The benefit of an omni-tomography scanner can be seen in this scenario, where a single scanner installed in a room can be used to perform most of these scans in a single visit, with periodic follow-up scans over the course of treatment. Furthermore, data collected from multiple imaging modalities on an omni-tomography scanner can be seamlessly integrated, packaged, and presented to clinicians for analysis and diagnosis. This can allow better management and tracking of the progression of the cancer.

With the current increasing trends in modality usage, we can expect that hospitals and clinics will be interested in a scanner that can support all imaging modalities, while at the same time providing more information per scan. The benefit of an omni-tomography scanner may therefore be quantified in terms of the number of image modalities per patient scan. A dual modality scanner may provide 2 image modalities/patient scan compared to an omni-tomography scanner that provides 4 to 5 imaging modalities/patient scan.

*Table II. Cost estimation for the proposed omni-tomography (also referred to as multi-tomography) system.*

|  | *Component* | *Manufacturer* | *Elemental Cost ($USD)* | *Commercial Scanner Cost ($USD)* | *Omni-tomography Elemental Cost ($USD)* |
|---|---|---|---|---|---|
| **Magnetic Resonance Imaging (MRI)** | Permanent magnet, electronics, gradient/RF coils | Wenzhuo Light Industrial | $50,000–$150,000 | Open MRI $200,000–$1,000,000 | $150,000 (Alibaba) |
| **Computed Tomography (CT)** | Tube, electronics, and detector | Varian, Radiation Monitoring Devices Inc. | $25,000–$50,000 | $1,300,000 | $30,219.28 (eBay) tube, electronics, and detector |
| **Positron Emission Tomography (PET)** | Block crystal and coincidence circuit | Radiation Monitoring Devices Inc. | $50/cm$^3$, 20,000 crystals per ring, Up to $1,000,000 | $1,300000 | High resolution, Low Cost PET Module, $256,200–$400,000 |
| **Single Photon Emission Computed Tomography (SPECT)** | CZT detector and electronics | Radiation Monitoring Devices Inc. | Up to $800,000 | $1,050000 | $400,000–$500,000 CZT systems (10 x10 x 2 cm$^3$) |
| **Ultrasound Imaging (US)** | US transducers and electronics | Boston Scientific | $1,500 per transducer | $10,000–$40,000 | $30,000 |
| **Optical Imaging** | Near infrared photon-counting camera, x-ray fluorescence camera | Princeton Scientific, Amptek | $5,000-$40,000 per camera | $500,000 | $100,000 |
| *Estimated Total Cost* | | | | | **About $1,500,000** |

## III. Interior Reconstruction

Interior tomography is both theoretical basis and enabling technology for omni-tomography or multi-tomography, because it or its variants (such as approximations) must be used to deal with truncated data. In this section, we present representative examples to establish the feasibility of interior reconstruction in the context of omni-tomography. Note that some modalities for omni-tomography can have less truncation in the data acquisition process than others.

### A. Interior CT

The long-standing interior problem is to reconstruct an interior ROI only from projection data associated with lines through the ROI. Since 2007, interior tomography theory and methods have been developed to solve this interior problem in a theoretically exact and stable fashion under practical conditions [50, 53, 57-58]. In the compressive sensing (CS) framework, we proved that an interior ROI can be exactly reconstructed via the total variation (TV) / high-order TV (HOT) minimization if features to be reconstructed is piecewise constant/polynomial [51-52, 59-60]. Most importantly, numerical and experimental results using interior tomography have been very encouraging.

**Imaging Model** – In a fan-beam / cone-beam CT geometry, the forward and back projection operations for image reconstruction usually assume a linear integral from the detector center to the source point. Let $f(\mathbf{x})$ be a 2D compactly supported function. The x-ray projection is modeled by

$$P(\mathbf{a},\boldsymbol{\beta}) = \int_0^\infty f(\mathbf{a}+t\boldsymbol{\beta})dt, \tag{III.A.1}$$

where $\mathbf{a} \in \mathbb{R}^2$ represents a source position, and $\boldsymbol{\beta} \in \mathbb{S}$ a 2D unit vector. Here we introduce a more realistic imaging model [61]. Let $\Omega_s \subset \mathbb{R}^2$ be a compact support of a source spot. As shown in **Figure III.A.1**, the two detector elemental border points and $\mathbf{a} \in \Omega_s$ define a narrow fan-beam of angle $\gamma_\mathbf{a}$. A unit vector from $\mathbf{a}$ to a point in the detector aperture is $\boldsymbol{\beta}(\theta,\mathbf{a})$, $\theta \in [0,\gamma_\mathbf{a}]$. Assuming that the x-ray flux along any direction from $\mathbf{a} \in \Omega_s$ is $I_0$, the signal detected by the detector element can be computed according to Beer's law:

$$I(\mathbf{a},\theta) = I_0 e^{-\int_0^\infty f(\mathbf{a}+t\boldsymbol{\beta}(\theta,\mathbf{a}))dt}, \tag{III.A.2}$$

which is equivalent to Eq. (III.B.1) after a logarithm operation $\ln(I_0/I(\mathbf{a},\theta))$. Considering the finite detector and source sizes, the total number of received photons becomes

$$I = \int_{\Omega_s} \int_0^{\gamma_\mathbf{a}} I(\mathbf{a},\theta) d\theta d\mathbf{a} = I_0 \int_{\Omega_s} \int_0^{\gamma_\mathbf{a}} e^{-\int_0^\infty f(\mathbf{a}+t\boldsymbol{\beta}(\theta,\mathbf{a}))dt} d\theta d\mathbf{a}, \tag{III.A.3}$$

where $d\mathbf{a}$ represents an area differential of the focal spot. After appropriate approximation, it has been shown that Eq.(III.A.3) can be re-expressed as

$$\overline{\overline{p}} = \frac{\int_{\Omega_s} \frac{1}{\gamma_\mathbf{a}} \int_{\Omega_b} f(\mathbf{x}) \frac{d\mathbf{x}}{\|\mathbf{x}-\mathbf{a}\|} d\mathbf{a}}{\int_{\Omega_s} d\mathbf{a}}, \tag{III.A.4}$$

where $\Omega_b$ denotes the narrow fan-beam region, and $d\mathbf{x}$ the corresponding differential [61].

For digital image reconstruction, $f(\mathbf{x})$ is discretized as a discrete image $\mathbf{f} = (f_{i,j}) \in \mathbb{R}^{N_I} \times \mathbb{R}^{N_J}$, $1 \le i \le N_I$ and $1 \le j \le N_J$. Let us define

$$f_n = f_{i,j}, \quad n = (i-1) \times N_J + j, \tag{III.A.5}$$

$1 \leq n \leq N$ and $N = N_J \times N_J$, and we have a vector $\mathbf{f} = [f_1, f_2, ..., f_N]^T \in \mathbb{R}^N$. We can use either $f_{i,j}$ or $f_n$ to represent the image. Assume that the finite focal spot $\Omega_s$ be discretized as $\mathbf{a}_q$, $1 \leq q \leq Q$. Let $\mathbf{p}^q = [p_1^q, p_2^q, ..., p_M^q]^T \in \mathbb{R}^M$ is measured data associated with the x-ray source sub-region $\mathbf{a}_q$ and all the detector elements, where $M$ is the product of the number of projections and the number of detector elements. By Eq. (III.B.4), we have the following discrete linear system

$$\mathbf{p} = \frac{1}{Q}\sum_{q=1}^{Q}\mathbf{p}^q = \frac{1}{Q}\sum_{q=1}^{Q}\mathbf{B}^q\mathbf{f} = \left(\frac{1}{Q}\sum_{q=1}^{Q}\mathbf{B}^q\right)\mathbf{f} = \mathbf{Bf}, \quad (III.A.6)$$

where $\mathbf{B}^q = (B_{m,n}^q) \in \mathbb{R}^M \times \mathbb{R}^N$ is the linear measurement matrix for the x-ray source $\mathbf{a}_q$. As shown in **Figure III.A.2**, the $n^{th}$ pixel is viewed as a rectangular region with a constant value $f_n$, the $m^{th}$ measured datum $p_m^q$ is viewed as an integral of those areas of pixels that are partially covered by a narrow fan beam from the source sub-region $\mathbf{a}_q$ to a detector element and weighted by the corresponding x-ray linear attenuation coefficients and fan-arc lengths. Thus, the component $B_{m,n}^q$ in Eq. (III.A.6) can be expressed as

$$B_{m,n}^q = \frac{S_{m,n}^q}{L_{m,n}^q}, \quad (III.A.7)$$

where $S_{m,n}^q$ denotes the intersection area between the $n^{th}$ pixel and the $m^{th}$ fan-beam path, and $L_{m,n}^q$ can be approximately computed as the product of the narrow fan-beam angle $\gamma_m^q$ and the distance from the $n^{th}$ pixel center to the x-ray source $\mathbf{a}_q$, which can be viewed as fan-arc length of the narrow fan-beam through the $n^{th}$ pixel center.

**Image Reconstruction –** The CS approach has recently been used in multiple biomedical imaging applications. With CS, high-quality signals and images can be reconstructed from far fewer data/measurements than required by the Nyquist sampling theory [62-63]. The main idea of CS is that most signals are sparse (i.e. a majority of their coefficients are close or equal to zero) when analyzed in an appropriate domain. Hence, CS depends on a sparsifying transform. In CS-based image reconstruction, commonly used sparsifying transforms are: discrete gradient transforms and wavelet transforms. The discrete gradient transform was successfully utilized in CT reconstruction [51, 64], given that the x-ray attenuation coefficient usually varies mildly within organs and large variations are usually found across boundary borders.

A CS solution to Eq. (III.A.7) can be expressed as

$$\hat{\mathbf{f}} = \arg\min_{\mathbf{f}} \|\mathbf{p} - \mathbf{Bf}\|^2 + \beta TV(\mathbf{f}), \quad (III.A.8)$$

where the term $\|\mathbf{p} - \mathbf{Bf}\|^2$ is for the data discrepancy, $TV(\mathbf{f})$ is for the image sparsity and $\beta$ is a balancing parameter. We developed an iterative algorithm to solve Eq. (III.A.8.) [51, 65], which consists of two major steps: (1) the ordered-subset simultaneous algebraic reconstruction technique (OS-SART) [66] is used to reconstruct an ROI image based on all the truncated projection data; (2) the TV is minimized in a soft-threshold filtering framework with a pseudo-inverse discrete gradient transform [65]. These steps are performed in an alternating manner. To accelerate the convergence, we employed the projected gradient method [67] to determine an optimal filtering threshold and implemented a fast iterative version [68].

**Numerical Simulation –** According to our current omni-tomography design (**Figure II.A.1**), we performed numerical simulations using a circular scanning locus of radius 45.0 cm and fan-beam geometry. An equi-space detector was placed opposite to the x-ray source with a source-to-detector distance 87.5 cm), and positioned perpendicular to the direction from the source to the system origin. The detector array consisted of 384 square elements, each of which was 0.776 mm in length, giving a total detector width of 29.7 cm. The detector size was chosen according to the commercial flat-panel detector PaxSca4030 in the 4x4 binning mode. We assumed to have an ideal point x-ray source, which means $Q = 1$ in Eq. (III.A.6). This configuration can cover an interior ROI of diameter 15.1 cm. A 1000 x 1000 patient chest image phantom was used, covering an area of 40X40 cm$^2$. The heart was centered at the system origin. For a full-scan, we acquired 1160 equi-angular projections, which was within the range used in commercial CT scanners. **Figure III.A.3** presents the interior reconstruction after 15 iterations, which shows that there is no significant difference between the interior reconstruction and the original phantom image.

## B. Interior MRI

For our proposed omni-tomography scanner, conventional MRI magnets do not seem a viable option because of their spatial bulkiness. Our alternative strategy is to create a local magnetic background field (main field) such as using permanent magnets. This local magnetic field can be either homogeneous or inhomogeneous. It is possible to create a homogeneous local magnetic field using relatively small permanent magnets, as shown in **Figure II.B.1**. So long as the homogeneity of the magnetic field holds over an ROI, classical MRI approaches (and correction methods) can be adapted for local MRI (see a recent review paper [69] and references therein, as well as the discussion below on how to deal with an inhomogeneous magnetic background field).

Alternatively, we propose here a novel approach based on an inhomogeneous local magnetic background field for local MRI, which is referred to as interior MRI. The key idea is to take advantage of surfaces with equal magnetic field strength, or "level sets", defined by the inhomogeneous background field for spatial localization. This approach accommodates a certain degree of field non-homogeneity, and is fundamentally different from the conventional MRI where we start with a homogeneous main magnetic field. Thus, it reduces the size (and possibly also cost) of the main magnet and is particularly attractive for omni-tomography or multi-tomography.

**Imaging Model** – In the case of interior MRI, without loss of generality, an ROI $D$ is represented by the union of level sets of the main field $B_0$ along the z-direction (in the same spirit we can deal with the main field $B_0$ whose magnetic lines are not necessarily parallel to the z-direction), i.e.,

$$D = \bigcup_c \{\vec{x} : B_0(\vec{x}) = c\}, \quad c \in [B_{\min}, B_{\max}], \qquad (III.B.1)$$

where $\vec{x}$ denotes a spatial point. In other words, the inhomogeneous magnetic field $B_0$ offers a natural spatial decomposition into level sets, which are constant intensity sub-regions or iso-regions. As the first step for spatial localization, we can tune the radiofrequency (RF) pulse to excite any iso-region, as shown in **Figure III.B.1(a)**. Then, standard x-, y- and z- linear gradient fields can be used to further localize each iso-region into desired voxels. However, this may require a long data acquisition time when the number of iso-regions is large, since each iso-region needs a 3D scan. To achieve a sampling efficiency that is equivalent to that of the conventional MRI scan, we can use a compressive sampling scheme, which will be illustrated later (in 2D) when the iso-regions are circular.

In practice, an inhomogeneous field can be well controlled. As a simple example, **Figure III.B.2** shows a magnetic field consisting of cylindrical iso-surfaces with a field strength decreasing away from the z axis and a field direction parallel to the z axis. For such an inhomogeneous magnetic field, 2D slices (e.g. transverse or z-slices) can be selected by modulating the z-gradient field appropriately. For example, we can keep a constant z-gradient field strength at one and only one z-level and vary the z-gradient field quickly outside the given z-level, as shown in **Figure III.B.1(b)**. Such a 2D slice intersects iso-regions and generates magnetic resonance (MR) data from the given z-slice only. In other words, by keeping the z-gradient constant at a given longitudinal location, the desirable iso-regions at the z-slice can be constructively excited, while by varying the z-gradient rapidly outside that longitudinal location, the iso-regions off the z-slice are incoherently excited for a short time, and their excitation effect can be negligible, as illustrated in **Figure III.B.2**. In this way, we can select any 2D slice. Note that this principle can be extended to a general smooth magnetic field.

As mentioned above, in interior MRI we utilize the level sets naturally defined by an inhomogeneous main field for spatial localization, and acquire MR data aided by standard linear gradient fields. During data acquisition, the MR signal is generated with dynamic radiofrequency (RF) pulses and can be recorded for every iso-region.

For higher acquisition efficiency, we propose the following compressive sensing scheme. Without loss of generality, let us consider interior MRI of a thin 2D x-y slice under the main field $B_0$ along the z direction that is radially symmetric in the x-y plane, i.e.,

$$B_0(r,\theta,z) = B_0(r^\alpha), \tag{III.B.2}$$

and linear x and y gradients with the slope $G_x$ and $G_y$ respectively. During data acquisition, we excite each iso-curve with the RF pulse that has the same excitation and demodulation frequency as the resonance frequency for the corresponding level set. That is, to excite the iso-curve with its radius $r=r_0$, the resonance frequency should be

$$\omega_0(r,\theta) = \gamma B_0 \delta(r - r_0), \tag{III.B.3}$$

the demodulated signal, assuming the uniformity of the receiver coil, is proportional to

$$s(t) = \int \omega_0 \rho e^{i\gamma(xG_x + yG_y)t} dx dy, \tag{III.B.4}$$

where $\rho$ represents a 2D MR image to be reconstructed.

**Compressive Sensing** – After the Fourier space is filled, by varying the y gradient field as the phase-encoding gradient, and acquiring the signal with the x gradient field as the frequency-encoding gradient, the iso-curve of the image at $r=r_0$ can be reconstructed using the inverse Fourier Transform. Continuing in this fashion, we can recover all iso-curves and therefore reconstruct the entire image. However, this requires excessive acquisition time, on the order of a 3D scan, since it takes one 2D scan for each iso-curve.

Fortunately, we have devised the following data acquisition scheme which should take a similar amount of time as a conventional 2D MRI scan. Before the compressive sensing acquisition, we randomize the gradient field orientation indexes and the iso-curve indexes, respectively, and then sequentially pair the field orientations with the iso-curves, as illustrated in **Figure III.B.1(b)**. According to this mutated combination, we sample each iso-curve under only one gradient orientation, in contrast with all the phase-encoding steps described previously. Now, we represent gradient fields ($G_x$, $G_y$) by ($G$, $\theta_0$),

$$G_x = G\cos\theta_0, \quad G_y = G\sin\theta_0, \quad \theta_0 \in [0,\pi], \tag{III.B.5}$$

where the orientation angle $\theta_0$ is random but fixed for a given iso-curve. The reason for randomization is as follows: for any iso-curve, since the data are acquired by varying $G$ for some fixed angle $\theta_0$, the spatial localization is non-unique on the circle; yet no continuous function defined on the circle can resolve the non-uniqueness. To overcome this ambiguity, we start with the random selection of $\theta_0$ for each iso-curve, and then perform image reconstruction from data associated with all the iso-curves subject to the underlying sparsity in a proper transform domain, such as the total variation (TV). With angular randomization and compressive sensing techniques, the non-unique spatial locations on the iso-curve can now be separated to satisfy certain image smoothness conditions. As a result, the entire image can be accurately reconstructed. As in **Figure III.B.3**, the image is assumed to consist of 4 iso-curves. Instead of taking all phase-encoding steps for each iso-curve, we just use one step for the iso-curve. This acquisition is random in terms of the gradient field orientation, which is from a pre-specified set of admissible angles. Then, the image can be reconstructed through a $L_1$-type minimization.

**Image Reconstruction** – On the discrete level, we can rewrite the MR signal equation in polar coordinates for the level set with $r=r_0$ from Eqs. (III.B.2)-(III.B.5) as

$$s(k_\theta) = r_0 B_0\big|_{r=r_0} \int \rho(r_0,\theta) e^{ik_\theta[2\pi r_0 \cos(\theta-\theta_0)]} d\theta, \tag{III.B.6}$$

where $k_\theta$ is defined as

$$k_\theta = \frac{\gamma G}{2\pi} t. \tag{III.B.7}$$

Then, on the discretized Cartesian grid, an image $\rho$ satisfying Eq. (III.B.6) can be represented by

$$A\rho = s, \tag{III.B.8}$$

where $A$ is the system matrix discretized from Eq. (III.B.6) for all iso-curves, incorporating the interpolation effect from polar to Cartesian coordinates with the polar grid oversampled near the origin to maintain the integral accuracy.

Under the least-square data fidelity and the TV regularization, interior MRI can be formulated as

$$\rho = \arg\min_{\rho} \|A\rho - s\|_2^2 + \lambda \|\rho\|_{TV}, \tag{III.B.9}$$

where $\lambda$ is the regularization parameter that balances data fidelity and image smoothness. Here we apply the split Bregman method [70] as an efficient solver of Eq. (III.B.9). To demonstrate the merits of the compressive sensing technique, we evaluate Eq. (III.B.9) against the classic L2 regularization, which is similar to the Tikhonov regularization or the maximum-likelihood method

$$\rho = \arg\min_{\rho} \|A\rho - s\|_2^2 + \lambda \|\rho\|_2^2. \tag{III.B.10}$$

**Numerical Simulation –** We performed the numerical simulation with an MRI cardiac image as the phantom. We compared our L1 method with the classic L2 method to demonstrate the need for compressive sensing. The results clearly indicated that the L1 method outperformed the L2 method in terms of contrast and signal to noise ratio (SNR). To further demonstrate the potential of our L 1 method, we under-sampled the data space by a factor of 50%, 25%, 12.5% respectively, and our method still reconstructed well. Moreover, we showcased ROI-based reconstruction, pertinent to omni-tomography, in which MR data was only acquired within an ROI, as shown in **Figure III.B.4**. The results showed that the proposed technique enabled accurate interior image reconstruction based on a local inhomogeneous main field.

One possible drawback for the application of an inhomogeneous magnetic field is its influence on transverse relaxation time (T2*) which is, to a large degree, characterized by the $B_0$ non-homogeneity. This can be mitigated by reducing the strength of the measurement field and using a proper sequence such as spin-echo. New imaging techniques such as UTE and SWIFT [71-72] that image objects consisting of spins with extremely short T2 can be adapted for our case, with negligible time between excitation and signal acquisition. It is emphasized that what we have discussed serves mainly as a proof of concept. The field of MRI has accumulated a huge amount of results, most of which are based on a homogeneous magnetic background field. We hope that the idea on MRI based on an inhomogeneous magnetic background field will bring new results that can be practically used in the near future.

The inhomogeneous magnetic field based MR model, compressive sampling and reconstruction techniques for interior MRI can be extended to other types of 2D and 3D magnetic field level sets and other MR sequences. For example, to accelerate the data acquisition process, multiple level sets can be simultaneously excited with RF pulses of different frequencies, and multi-frequency signals can be received through multi-channels. As another example, the magnetic force lines of an inhomogeneous magnetic background field can be assumed to be curvilinear instead of being straight, requiring more sophisticated sequences and reconstruction formalisms.

## C. Interior SPECT

Inspired by the development of interior tomography for CT, interior SPECT techniques were also developed assuming a constant attenuation background (further work in progress). SPECT is used to reconstruct a radionuclide (radiotracer) source distribution from externally measured gamma ray photons. A gamma camera is used to acquire multiple 1D/2D projections from various angles. Then, a CT-like algorithm is applied for image reconstruction. Based on interior tomography results [50, 53, 57-58], we proved in 2008 that theoretically exact interior SPECT is feasible from uniformly attenuated local projection data, aided by prior knowledge of a sub-region in the ROI [73]. With CS techniques [51-52, 59-60], we further proved that if an ROI is piecewise polynomial, then it can be uniquely reconstructed from truncated SPECT data directly through the ROI [74].

**Imaging Model –** Let $f(\mathbf{x})$ be a 2D smooth distribution function on a compact support $\Omega$ with $\mathbf{x} = (x, y) \in \Omega$. In a parallel-beam geometry illustrated in **Figure III.C.1**, SPECT projections of $f(\mathbf{x})$ can be modeled as an attenuated Radon transform [75]

$$P_o(\theta,s) = \int_{-\infty}^{\infty} f(s\boldsymbol{\theta}+t\boldsymbol{\theta}^{\perp})e^{-\int_{t}^{\infty}\mu(s\boldsymbol{\theta}+\tilde{t}\boldsymbol{\theta}^{\perp})d\tilde{t}} dt, \qquad (\text{III.C.1})$$

where the subscript "o" denotes original projection data, $\boldsymbol{\theta}=(\cos\theta,\sin(\theta))$, $\boldsymbol{\theta}^{\perp}=(-\sin\theta,\cos(\theta))$, and $\mu(\mathbf{x})$ the attenuation coefficient map on the whole support. In practice, the attenuation map can be approximated as a uniform distribution

$$\mu(\mathbf{x}) = \begin{cases} \mu_0 & \mathbf{x}\in\Omega \\ 0 & \mathbf{x}\notin\Omega \end{cases}, \qquad (\text{III.C.2})$$

where $\mu_0$ is a constant. Since the object function is compactly supported, we can determine the length of the intersection between the support $\Omega$ and the integral line for $P_o(\theta,s)$. Without loss of generality, we denote this length as $t_{\max}(\theta,s)$, and Eq. (III.C.1) becomes

$$P_o(\theta,s) = e^{-\mu_0 t_{\max}(\theta,s)} \int_{-\infty}^{\infty} f(s\boldsymbol{\theta}+t\boldsymbol{\theta}^{\perp})e^{\mu_0 t} dt. \qquad (\text{III.C.3})$$

Let us assume that the compact support $\Omega$ and the constant coefficient $\mu_0$ are known. By multiplying a weighting factor $e^{\mu_0 t_{\max}(\theta,s)}$, the projection model for SPECT is reduced to

$$P_w(\theta,s) = P_o(\theta,s)e^{\mu_0 t_{\max}(\theta,s)} = \int_{-\infty}^{\infty} f(s\boldsymbol{\theta}+t\boldsymbol{\theta}^{\perp})e^{\mu_0 t} dt, \qquad (\text{III.C.4})$$

where the subscript "w" indicates weighted projection data. In this context, CT may be regarded as a special case of SPECT. However, generally CT reconstruction techniques cannot be directly used for SPECT.

The discrete CT model Eq. (III.A.6) with $Q=1$ can be extended to the SPECT case:

$$\mathbf{P}_w = \tilde{\mathbf{B}}\mathbf{f}, \qquad (\text{III.C.5})$$

where $\tilde{\mathbf{B}} = (B_{m,n}w_{m,n}^{\mu_0})$, and $w_{m,n}^{\mu_0}$ is the corresponding discrete term $e^{\mu_0 t}$ in Eq. (III.C.4).

**Image Reconstruction** – Suppose that $f(\mathbf{x})$ is a piecewise n-th ($n\geq 1$) order polynomial function in $\Omega$ [74]; that is, $\Omega$ can be decomposed into finitely many sub-domains $\{\Omega_i\}_{i=1}^{I_r}$ such that

$$f(\mathbf{x}) = f_i(\mathbf{x}), \text{ for } \mathbf{x}\in\Omega_i, \ 1\leq i\leq I_r, \qquad (\text{III.C.6})$$

where $f_i(\mathbf{x})$ is a n-th order polynomial function, and each sub-domain $\Omega_i$ is adjacent to its neighboring sub-domains $\Omega_j$ with piecewise smooth boundaries $\Gamma_{i,j}$, $j\in N_i$. Then, the high-order total variation can be defined as [74]:

$$\text{HOT}_{n+1}(f) = \sum_{i=1}^{I_r}\sum_{j>i,j\in N_i}\int_{\Gamma_{i,j}}|f_i - f_j|ds + \int_{\Omega}\min\left\{\sqrt{\sum_{r=0}^{n+1}\left(\frac{\partial^{n+1}f}{\partial x^r \partial y^{n+1-r}}\right)^2}, \sqrt{\left(\frac{\partial f}{\partial x}\right)^2+\left(\frac{\partial f}{\partial y}\right)^2}\right\}d\mathbf{x}, \qquad (\text{III.C.7})$$

where the second term is a Lebesgue integral. Here we used the following approximation [74]:

$$\text{HOT}_2^{\text{dis}} = \sum_{u,v}\sqrt{(D_{11}(u,v))^2 + (D_{12}(u,v))^2 + (D_{22}(u,v))^2}, \qquad (\text{III.C.8})$$

where $D_{11}(u,v) = f_{u+1,v} + f_{u-1,v} - 2f_{u,v}$, $D_{12}(u,v) = (f_{u+1,v+1} + f_{u-1,v-1} - f_{u-1,v+1} - f_{u+1,v-1})/4$, and $D_{22}(u,v) = f_{u,v+1} + f_{u,v-1} - 2f_{u,v}$ are the second-order finite differences.

With $\text{HOT}_2$, the solution of interior SPECT can be found as

$$\hat{\mathbf{f}} = \arg\min_{\mathbf{f}} \|\mathbf{P}_w - \tilde{\mathbf{B}}\mathbf{f}\|^2 + \beta\text{HOT}_2(\mathbf{f}). \qquad (\text{III.C.9})$$

Comparing Eqs. (III.C.9) with (III.A.8), we see the same computational structure. Therefore, we implemented an interior SPECT algorithm based on the interior CT algorithm [74]. The major difference between the two algorithms is computation of the steepest gradient direction.

**Numerical Simulation** – As shown in **Figure III.C.2**, we downloaded a SPECT cardiac perfusion image from the Internet and modified it into a realistic 128X128 image phantom, covering an area of 128X128 mm$^2$. It represents a radionuclide distribution in a human heart. In our simulation, we assumed a constant attenuating background $\mu_0 = 0.15$ cm$^{-1}$ on a compact support of a standard patient size. We used an equi-spatial detector array consisting of 78 detector elements, each of which was 1.0 mm in length. For a full-scan, we acquired 128 equi-angular projections. Clearly, the realistic image phantom does not satisfy the piecewise polynomial model in the case of n=1. To improve the stability of interior SPECT, two additional constraints were incorporated into the OS-SART loop in the projection-onto-convex-sets (POCS) framework: (1) non-negativity, which means that the radionuclide distribution should not be negative, therefore we made negative values zero during the iterative process; (2) compactness, which means that the radionuclide distribution should be inside the human body, and therefore we iteratively made the pixels outside a specified body-contour zero. It can be seen from **Figure III.C.2** that interior SPECT produced excellent results, even if the piecewise polynomial model is not exactly satisfied.

### D. Interior X-ray Fluorescence CT

X-ray fluorescence techniques are widely used for elemental analysis [76], and offer excellent sensitivity to trace elements down to the picogram level. Nanoparticles are particularly useful as imaging probes because of their unique and non-toxic physicochemical properties[77]. For example, nanoparticles can leak into the tumor via blood circulation and accumulate within it. X-ray fluorescence computed tomography (XFCT) can be used for 3D mapping of nanoparticles inside a human or animal subject. As shown in **Figure III.D.1**, most XFCT studies uses a pencil-beam of x-rays to illuminate an object and cause emission of fluorescent x-rays from a variety of elements. The fluorescent x-rays are collected by a sensitive x-ray spectrometer, and can be used to identify various elements of interest. Also, the object is scanned and rotated in the first-generation CT geometry to obtain sufficient data for reconstruction of the elemental distributions [76].

**Imaging Model** – When an x-ray beam travels through a tissue, it undergoes photoelectric interactions with inner shell electrons of atoms. The x-ray intensity distribution on the primary path can be computed by Beer's law:

$$f(\alpha,s,t) = I_0 \exp\left(-\int_{-\infty}^{t} \mu_t(s,t')dt'\right) \qquad \text{(III.D.1)}$$

where $I_0$ is the x-ray source intensity, $\mu_t$ the attenuation coefficient which can be obtained from CT, and $(s,t)$ are rotation coordinates of $(x,y)$ $s = x\cos\alpha + y\sin\alpha$, $t = -x\sin\alpha + y\cos\alpha$. When x-ray photons interact with matter, the involved atoms isotropically emit characteristic x-rays with intensity proportional to the product of the absorbed x-ray flux rate and the fluorescent x-ray yield at a point [78]:

$$g(\alpha,s,t) = \varepsilon\mu_{ph} f(\alpha,s,t) \int_{\theta_1}^{\theta_2} \exp\left[-\int_{0}^{\infty} \mu_F(s - b\sin\beta, s + b\cos\beta)db\right]d\beta \qquad \text{(III.D.2)}$$

where $\mu_{ph}$ is the photoelectric linear attenuation coefficient, $\varepsilon$ the fluorescent x-ray yield, $\theta_1$ and $\theta_2$ define the angular aperture of the x-ray fluorescence detector, as shown in **Figure III.D.1**. The total flux rate of the fluorescent x-ray reaching the fluorescence detector is obtained by integration:

$$I(\alpha,s) = \int_{-\infty}^{\infty} f(\alpha,s,t)g(\alpha,s,t)\rho(s,t)dt \qquad \text{(III.D.3)}$$

where $\rho(s,t)$ is the concentration of nanoparticles.

**Image Reconstruction –** XFCT reconstruction is a linear inverse problem when the linear attenuation coefficient distributions of the object are known at the energies of the incident and fluorescent x-rays. Hence, Eq. (III.D.3) can be discretized into the matrix format using numerical integration methods:

$$\mathbf{AX} = \mathbf{B}, \quad (\text{III.D.4})$$

where **X** is an discretized image in terms of a nanoparticle density distribution, **B** represents measured fluorescent signals, and **A** the system matrix from Eq. (III.D.3). Using the CS principle, we incorporated the L1 norm regularization for the sparse representation of the nanoparticle density distribution subject to the data constraint. Finally, the solution can be obtained as follows:

$$\hat{\mathbf{X}} = \arg\min_{\mathbf{X}} \left\{ \|\mathbf{AX} - \mathbf{B}\|^2 + \lambda \|\mathbf{TX}\|_1 \right\} \quad (\text{III.D.5})$$

where T represents a sparse transform, and λ a regularization parameter. Eq. (III.D.5) can be efficiently solved using a contemporary optimization method such as the Bregman iteration scheme [79].

**Numerical Simulation –** We performed a numerical test of interior x-ray fluorescence computed tomography for a 2D patient slice. The phantom consisted of gold nanoparticles accumulated in four circular regions with radii of 3.0 mm, 7.0mm, 4mm and 5 mm and Gaussian-like concentration distributions of 3 µg/ml, 5 µg/ml, 3.5 µg/ml, and 4 µg/ml, respectively. Two additional clusters of nanoparticles were targeted in an ROI. An x-ray source of 110 keV and 30 mA was collimated into a pencil beam to irradiate the phantom. The image acquisition was repeated 100 times when the x-ray pencil beam was translated in a 0.1 mm increment for a given view to cover the ROI. The parallel-beam imaging geometry was rotated 50 times to span a 180° range evenly around the ROI. The projection data were generated according to Eq. (III.D.3), and Poisson noise was added to simulate realistic experimental conditions. The image was reconstructed using the shrinkage-thresholding algorithm based on Eq. (III.D.5) [79]. **Figure III.D.2** presents a typical interior XFCT reconstruction, which is in excellent agreement with the true image.

## IV. Discussions and Conclusion

### A. Trend of Hybrid Imaging

Tomography is widely used for preclinical and clinical imaging to characterize morphology, and to a limited extent biological function (e.g. physiology). Given current technology, we are forced to accept several intrinsic limitations of tomography, especially the necessity to acquire, reconstruct and analyze data obtained sequentially on the same subject with or without explicit superimposition of the results. This separation in time impairs the ability to decipher and understand biological functions, as they are certainly dynamic processes (especially relative to morphological changes). For example, a cardiac infarct commonly begins with decreased perfusion, then tissue hypoxia and eventually cell death; these stages exist in a continuum that can evolve over a short time frame relative to the time needed to acquire multimodality imaging data.

If technology could be developed to simultaneously image physiome and other dynamic complexities with multiple modalities, biological processes which evolve rapidly may become transparent. Such processes have temporal evolution at many intervals, including mille- or micro-seconds, seconds, minutes, days or longer, and may or may not be reversible. A single session imaging time frame is important to image processes such as: ischemia, drug interactions, radiation effects, apoptosis, and many others. To some extent, this has already been accomplished with PET-CT and MRI-PET systems. Although many of these multimodality imaging systems still acquire data sequentially, the delay in data acquisition is improved relative to single-modality predecessors.

Now that a formal mechanism for unifying structural and functional data is attracting interest, especially with the Physiome project, the need to simultaneously acquire and unify multimodal images has become more important than ever before. There are critical and immediate needs to remove the limitations inherent in today's tomographic imaging approaches to the extent that complex dynamic biological processes can be studied *in vivo* and in real-time using multiple modalities.

In a recent review article entitled "*Multimodality Imaging: Beyond PET/CT and SPECT/CT*" [34], Dr. Simon Cherry wrote that "*Multimodality imaging with PET/CT and SPECT/CT has become commonplace in clinical practice and in preclinical and basic medical research. Do other combinations of imaging modalities have a similar potential to impact medical science and clinical medicine? The combination of PET or SPECT with MRI is an area of active research at the present time, while other, perhaps less obvious combinations, including CT/MR and PET/optical also are being studied.*"

Indeed, since the seminal paper by Dr. Bruce Hasegawa [80], subsequent work on combining emission and transmission tomography has led to powerful hybrid imaging capabilities that have had major impacts on biomedical applications. The popularity of these hybrid modalities is a testament to the important value of tight association between anatomical information from CT and functional information from SPECT/PET. The first PET-CT system was developed in the late 1990's [81], and has been subsequently applied in the biomedical field to study various forms of cancer (e.g. brain, lung, thyroid, lymph, head and neck cancer), along with infection, inflammation and others. Several commercial PET-MRI scanners have been introduced over the past few years and are demonstrating promising results [30-31]. The power of the PET-MRI system lies in the synergy between PET and MRI. PET is capable of quantifying myocardial blood flow, assessing viability and prognosis, and is sensitive to many biomarkers of inflammation, angiogenesis, nervous functions, and therapeutic genes or cells.MRI determines ventricular function and structure, and also delineates infarction using a chelated-gadolinium enhancement agent. Another advantage of PET-MRI is its 4D imaging ability that allows for better contrast enhancement of small lesions through motion correction. Other modality fusion systems include the bioluminescence tomography (BLT) and micro-CT system developed by Dr. Wang's group in 2002, optical and MRI systems, hybrid nuclear and optical imaging systems, x-ray and MRI systems, photoacoustic tomography scanners, and more [82-85].

In the aforementioned article [34], Dr. Cherry raised a question that could represent the thought processes of experts in the field: "*Is the fusion of PET and SPECT with CT the ultimate answer in multimodality imaging, or is it just the first example of a more general trend towards harnessing the complementary nature of the different modalities on integrated imaging platforms?*" Our answer is to target the tightest possible integration of all imaging modalities, and methodically fuse the richest relevant information available from each technology. We will reach this target by applying the latest insights from interior tomography and compressive sensing principles to design our omni-tomography system – A system that places the highest demands on the broadest array of known imaging hardware and systems engineering. The logic seems clear that since subsets of imaging modalities are synergistic, the integration of all imaging modalities as a whole should, in principle, add values above that of the individual sets. From this perspective, we can safely say that some form of omni-tomography is certain to be developed in the near future. This is consistent with past medical imaging innovations, in that a major technical advancement can always find useful biomedical applications.

Major technical obstacles to omni-tomographic systems have been the gantry space limitation, the associated technical difficulties and also high cost. There are a few basic principles that guided our thinking about the proposed omni-tomography system: (1) interior imaging can be accurately applied to most imaging modalities; (2) interior imaging will be relevant in a majority of imaging studies; (3) our grand fusion approach will be a cost effective solution when all or many imaging modalities must be used. An immediate advantage of interior imaging can be seen in how the problem of SPECT-MRI must be addressed. SPECT-MRI has two unique issues: interference to the magnetic field caused by a rotating camera head, and the induction of eddy currents in the camera head itself. Fortunately, interior SPECT reconstruction will effectively address these issues simultaneously. With a smaller SPECT camera, the electromagnetic interference will be reduced, and electromagnetic shielding design will be simplified. Eddy currents will diminish in significance with proper shielding and smaller detector heads.

It is underlined that, in a substantial sense, omni-tomography covers what is known today as "*modality fusion*". With flexibility at a clinician's discretion, according to a proper protocol, our omni-tomography scanner allows use of a single modality or multiple modalities in simultaneous combination on an ad-hoc basis. For a diagnosis/treatment where the patient requires multiple scans from various imaging modalities, omni-tomography could be a cost-effective way in terms of space utilization, equipment costs and patient throughput. In a sense, omni-tomography could be a versatile cost effective imaging platform, comparable to a fully-fledged medical imaging center. The costs for startup and implementation of an omni-tomography system could be significantly lower than a large-scale hospital imaging department. Additionally, an omni-tomography scanner can contain a regular global tomographic imager to provide a global reference volume. For example, a composite interior imaging gantry and a regular CT imager can be put together in a PET-CT-like configuration.

## B. Grand Fusion in Three Aspects

Being consistent with the omni-tomography or multi-tomography concept, we conceive three inter-related lines of grand fusion: first, an architectural fusion that merges all or many imaging modalities into a single gantry; second, a component fusion that packs all or many detectors into a single device or chip; and third, a methodical fusion for data processing and image reconstruction in a unified framework. In the previous sections, we have mostly discussed the architectural effort. Here we can briefly touch upon the others.

As far as the component fusion is concerned, there are already various detector systems that are designed on a chip. A micro-electromechanical-device (MEMS) US transducer has been built and characterized [86]. The piezoresistive and capacitive transducer types have been compared, showing that the piezoresistive transducer is superior to its capacitive counterpart. As for MRI, MRI on a chip has been explored [87], and a wireless MRI transmitter is feasible [88]. In addition to solid-state detectors for SPECT and CT, a CZT detector is also feasible for PET. This is made possible with special readout chips, such as the RENA3 IC chip. Also, digital systems are available for SPECT with CZT detectors and PET with crystal systems and avalanche photodiodes. The primary merit of a single detector chip for multimodality imaging is simultaneous data acquisition. The bulkiness is not attractive for the detectors that are based on scintillation crystals (PET/SPECT), transducers (US), or magnet coils (MRI), when available space is an issue. Using the multi-component technology, omni-tomographic imaging would be easier to implement. With a sufficiently large number of omni-tomographic imaging chains together, ultrafast or even instantaneous omni-tomographic imaging can be achieved. This point was touched upon in our previously proposed multi-source interior tomography project [89]. A second benefit of the component-level integration is cost reduction via mass production. Easier system maintenance could be another potential benefit.

Lastly, the methodology fusion can be a very exciting topic. Although various modalities use different physical mechanisms, the imaging target is the same human or animal subject. As such, our domain knowledge on this subject should be maximally utilized such as in the form of a population-based statistically-driven elastic atlas. In this regard, excellent work in the area of post-hoc image registration can be adapted for omni-tomographic reconstruction [90-92]. Then, constraints can be applied for correlation among information from component modalities and consistence with our prior knowledge. The newly developed CS -framework can be used to unify the data process and image reconstruction [11, 93-95].

## C. Enhancement with New Modalities

New medical imaging modalities are emerging, making omni-tomography both increasingly powerful and challenging. To meet this challenge, it appears that some form of interior imaging is unavoidable for simultaneous tomographic reconstructions using an increasing number of modalities. The power of omni-tomography lies with the increased relevance of *in vivo* information collected over an ROI in a single, simultaneous scan.

As mentioned in the first section, x-ray CT is moving towards multi-energy imaging. The enabling technology is spectroscopic (energy-sensitive) detectors. For example, Medipix is a series of spectroscopic detectors for x-ray micro-imaging, developed in collaboration with the European Organization for Nuclear Research (CERN). There are three current generations of this detector: Medipix1 has detector cells each measuring 170x170 µm$^2$; Medipix2 refines that to 55x55 µm$^2$, yet the performance is limited by the charge cloud effect; Medipix3 will have special circuitry to allow charge deposition analysis without spectral distortion, and support eight energy thresholds.

Recently, Pfeiffer *et al.* have made major progress towards x-ray phase-contrast and dark-field imaging with a conventional (non-coherent) x-ray tube [96]. The key idea uses phase stepping interferometry based on the Talbot effect, a periodic self-imaging phenomenon, to extract phase shift and small-angle scattering information. This approach relies on the use of three gratings to: define small imaging apertures and produce individually coherent secondary sources from a hospital-grade x-ray tube; constructively superimpose interference fringes at an appropriate distance; detect signals via a correlation process from wave-fronts distorted by an object; and form projective images from which tomography is feasible.

Although we have presented only one case study on interior reconstruction per selected imaging modality, some modalities have multiple variants to extract distinct yet complementary information. For example, MRI supports many imaging modes, including: dynamic contrast-enhanced MRI, diffusion-weighted MRI, functional MRI, pharmacologic MRI, MR elastography, MR temperature mapping, and MR spectroscopy, and more.

Recently, magnetic particle imaging (MPI) was developed as a complementary mode to MRI. MPI detects ferromagnetic nanoparticles when they are in field-free regions within a static magnetic field. The harmonic signals are induced by oscillating fields that only influence nanoparticles in field-free regions. Interior MPI can be developed in reference to the results in the interior tomography area assuming a known subregion, a piecewise polynomial model, or other forms of prior knowledge. The single-sided MPI design can be a candidate for omni-tomography in the preclinical setting. Furthermore, with an inhomogeneous magnetic background we could combine MRI and MPI with more advanced interior imaging methods to extract even more information at the functional, cellular and molecular levels.

In addition to those discussed above, there are certainly other potential modalities that could be integrated with omni-tomography, such as microwave tomography and endoscopic imaging. Furthermore, novel contrast agents and nanoparticles can be combined with omni-tomography to deliver critical information [97]. It is the ultimate goal of omni-tomography that all these imaging modes can be nonexclusively performed at the same time so that we can gain insight comprehensively, complementarily, and concurrently.

**D. Key to Systems Biology**

At the beginning of this paper, we mentioned the IUPS Physiome Project. Indeed, the omni-tomography, or multi-tomography, offers the best opportunities to observe well-registered tempo-spatial features in an unprecedented fashion and may reveal many unknown physiological, pathological, pharmaceutical and interventional interactions *in vivo*, and significantly improve the sensitivity and specificity of basic research, diagnosis and monitoring. This is the major hypothesis that cannot be tested without an omni-tomographic imaging system in place.

We believe that imaging technology development must target important biomedical problems. For example, biotechnology and bioinformatics have been developed to decode:
(1) Genomic/epigenetic signals at the DNA level associated with various forms of genomic signatures (e.g. single mutations, rare mutations, SNPs, copy number changes, indels, genomic instability index, etc.);
(2) Gene expressions at mRNA/miRNA/shRNA/gene/exon/splicing levels, and their various functions;
(3) Protein expression (protein complex, metabolic, etc.);
(4) Complex interactions and networks among these players on multi-scales.

Many of these studies are currently limited to cells (tissue samples, cell lines, etc.) and rarely go beyond studies *in vitro*, yet imaging *in vivo* should help facilitate translation to the organ/system/body level, where these advances should be applied!

It is our belief that there are several unique and critical features that biomedical imaging should provide: (a) <u>*in vivo*</u>: monitoring cells in their native tissue environment for the most accurate picture of their true behaviors; (b) <u>tomographic</u>: 3D mappings on multi-scales from cells, tissue, organs, systems, all the way to the whole body; (c) <u>dynamic</u>: temporal evolution of the system within broad ranges of time windows; (d) <u>comprehensive</u>: biological system interactions are extremely complicated and time-sensitive, thus each imaging modality should contribute some important facet about the system in a temporally-sensitive fashion. The combination of these four features is our sense of omni-tomography and its uniqueness for biomedical applications.

Systems biology should be a great driver for omni-tomography. Let us consider a biological subject as a system with inputs, circuitry with feedback loops, and outputs. Our ideal omni-tomographic imaging system, with significant help from imaging probes, can image some components of such a system in multiple dimensions (e.g. time, space, general characteristic dimensionalities). One can imagine the numerous ways this imaging system would be attractive and even indispensible to researchers. Perhaps, an omni-tomographic imaging platform to life scientists will be like the Large Hadron Collider to particle physics. In the following two subsections, let us examine some specific potential applications of omni-tomography.

### E. Insight into Major Diseases

Cardiovascular disease is the leading cause of death globally. The annual incidence of acute coronary syndrome (ACS) for individuals in Europe is estimated to be 1:80 to 1:170, while the incidence of chest pain leading to hospital admission for suspected ACS is much higher. Catheterized fluoroscopic angiography has been the original clinical standard for identifying stenotic lesions in the coronary arteries. Yet, other diagnostic imaging techniques have been researched and applied for this purpose, such as CT, MRI, PET, SPECT, intra-coronary ultrasound, and optical coherence tomography. The latest research emphasis is to improve understanding of pathobiology and genetics behind coronary artery diseases. There are critical and immediate needs for significantly better diagnostic performance despite the impressive advancement of imaging technology. For example, the ability to model high risk atherosclerotic lesions would be clinically invaluable. The goal would be to develop imaging methods and computational models to identify and predict high-risk lesions that may rupture, leading to coronary thrombosis and myocardial infarction. If our proposed system could be used to study preclinical infarction models, predict various outcomes, and promote clinical translations, it would be a tremendous tool for cardiovascular imaging.

Cancer is also a major category of highly complicated and fatal diseases. Roughly speaking, cancer causes over 10% of all human deaths worldwide, and the rate is rising in the developing world. Diagnosis of cancer requires *in vivo* and *in vitro* imaging examination, but is often discovered too late for some patients. The prognosis would be better if early screening was sensitive and specific. The treatment will be more effective if it could be guided by powerful imaging and sensing techniques, preferably with tumors clearly labeled. Oncological imaging already uses all the anatomical, functional, cellular and molecular modalities. For example, lung cancer imaging has utilized all major medical tomography modalities. CT defines air-tissue interfaces, detects nodules and tumors, and provides quantitatively accurate information. MRI measures airways and function with hyperpolarized helium-3. PET and SPECT improve lung cancer diagnosis and staging with radiotracers. Omni-tomography promises to be an ideal cancer imaging tool. It could have the ability to quantify malignancy, without the need for extra invasive procedures like biopsies. Also, a futuristic feature might use the omni-tomography data to build models and create a knowledge depository linked to genetic/epigenetic information and patient histories. Furthermore, such a depository could be leveraged with information from drug databases, human genomic databases, cellular structure and functional databases, and more. Ultimately, with all this quantitative information at hand, we could envision personalized medicine where an individuals' physiology and morphology data, obtained from a single scan in an omni-tomography system, could be algorithmically compared with a database to design a pharmaceutical therapy uniquely effective for that individual's biological makeup.

### F. Platform for Drug Development

Our proposed omni-tomographic system may be first developed for drug development and testing in animal models, with the aid of multimodality probes. Engineering imaging probes visible to multiple modalities is a hot topic in the field of bionanotechnology, and will have a profound impact on drug development and molecular medicine. Recently, exciting work has been reported on a combination of multiple nano-millimeter-sized components to facilitate multimodality imaging and even enable new imaging modes, such as with iron oxide and gold-coupled core-shell nanoparticles (NPs) with well-defined characteristics (shell thickness, core-shell separation, electronic, magnetic, optical, thermal, and acoustic).

We would like to underline the huge potential impact of combining our proposed omni-tomography system with multimodality probes, and especially in the field of pharmaceutical research. In this context, pharmacological interactions and dynamics can be noninvasively investigated from various aspects with a singular platform and probe. Furthermore, this combination could enable efficient repetition in longitudinal studies of animal models for these purposes. We feel that preclinical pharmaceutical research could be the first area to employ omni-tomography, as these studies are typically not subject to lengthy restrictive FDA approvals, and will set a stage for subsequent translational research and clinical trials. We hope that omni-tomography, given its depth, width, locality and flexibility, will outperform the current successes with modality fusion.

### G. Concluding Remarks

In the Second International Conference on Computational and Mathematical Biomedical Engineering, March 2011, Dr. Peter Hunter presented the opening lecture entitled "*A Bioengineer's View of Recent Development in VPH/ Physiome Projec*t". After Dr. Hunter's presentation, this paper's first author, Dr. Ge Wang gave a keynote speech on interior tomography, but changed the title from "*Interior Tomography*" to "*Interior Tomographysiome*". In this talk, he stated that "*tomographysiome goes beyond the modality fusion approach and integrates as many as possible tomographic information sources for studies on systems biomedicine, preclinical and clinical applications. Interior tomographysiome targets most informative, localized tomographic sampling for biomedical investigations, sensitive yet specific diagnosis, as well as personalized and even preventive interventions.*" Since then, our team has actively been working on this project, and is highly optimistic about this emerging area. To reflect the essential concept of tomographysiome, the two phrases "tomography" and "grand fusion" could be coined into tomogranphy [98-99]. While this term "tomogranphy" was used for much of our preliminary documentation, upon further reflection the authors felt the terms "omni-tomography" or "multi-tomography" best conveyed the concept of our synergistic multi-spectral, multi-temporal and multimodal approach. Other alternative names we considered include inter-tomography, tomographome, and so on.

In conclusion, we have systematically presented the concepts that we hope will push omni-tomography to the forefront of biomedical imaging for both pre-clinical and clinical applications. We briefly described a novel top-level design for an omni-tomographic scanner, and the techniques that enable its configuration. We selectively reported our preliminary results on interior reconstructions using interior MRI with an inhomogeneous magnetic field, interior CT, interior SPECT, and interior x-ray fluorescence tomography, all of which are in the compressive sensing framework. Finally, we discussed our understanding of the necessity, significance and potential of omni-tomography. While we feel very confident about our concepts and design for omni-tomography, we do not realistically expect instantaneous acceptance and implementation of our ideas. However, it is worth recalling the intense debate surrounding the first PET/CT scanner built over a decade ago, and certainly many of the same questions regarding efficacy and applicability will be raised in objection to omni-tomography. In light of the recent success PET/CT has enjoyed, perhaps omni-tomography will meet fewer objections and more quickly attract interest. Hence, we are determined to proceed, and welcome your collaboration.

**Acknowledgment**

This work was partially supported by NIH/NIBIB grant EB011785 and NIH/NHLBI grant HL098912. We would like to thank Drs. Tiange Zhuang, Erik Ritman, Erwei Bai, Robert Kraft, Craig Hamilton, Youngkyoo Jung, Wenbing Yun, Yantian Zhang, Alexander Katsevich, and others for helpful discussions.

**References**

1. Hunter, P., P. Robbins, and D. Noble, The IUPS human physiome project. Pflügers Archiv European Journal of Physiology, 2002. 445(1): p. 1-9.
2. Hunter, P., et al., Integration from proteins to organs: the IUPS Physiome Project. Mechanisms of Ageing and Development, 2005. 126(1): p. 187-192.
3. Hunter, P.J. and T.K. Borg, Integration from proteins to organs: the Physiome Project. Nat Rev Mol Cell Biol, 2003. 4(3): p. 237-243.
4. Rutman, A.M. and M.D. Kuo, Radiogenomics: Creating a link between molecular diagnostics and diagnostic imaging. European Journal of Radiology, 2009. 70(2): p. 232-241.
5. van Houten, V.M.M., et al., Molecular Assays for the Diagnosis of Minimal Residual Head-and-Neck Cancer: Methods, Reliability, Pitfalls, and Solutions. Clinical Cancer Research, 2000. 6(10): p. 3803-3816.
6. Ahn, P.H. and M.K. Garg, Positron Emission Tomography/Computed Tomography for Target Delineation in Head and Neck Cancers. Seminars in Nuclear Medicine, 2008. 38(2): p. 141-148.
7. Bockisch, A., et al., Hybrid Imaging by SPECT/CT and PET/CT: Proven Outcomes in Cancer Imaging. Seminars in Nuclear Medicine, 2009. 39(4): p. 276-289.
8. Delbeke, D., et al., Hybrid Imaging (SPECT/CT and PET/CT): Improving Therapeutic Decisions. Seminars in Nuclear Medicine, 2009. 39(5): p. 308-340.
9. Even-Sapir, E., Z. Keidar, and R. Bar-Shalom, Hybrid Imaging (SPECT/CT and PET/CT)--Improving the Diagnostic Accuracy of Functional/Metabolic and Anatomic Imaging. Seminars in Nuclear Medicine, 2009. 39(4): p. 264-275.
10. Kaufmann, P.A. and M.F. Di Carli, Hybrid SPECT/CT and PET/CT Imaging: The Next Step in Noninvasive Cardiac Imaging. Seminars in Nuclear Medicine, 2009. 39(5): p. 341-347.
11. Wang G, G.H., Zhang J, Weir V, Xu XC, Cong A, Bennett J, Medical Imaging, in iomedical Engineering Education and Advanced Bioengineering Learning: Interdisciplinary Concept, Z.O. Abu-Faraj, Editor. in press, IGI Global: Hershey, PA.
12. Petersilka, M., et al., Technical principles of dual source CT. European Journal of Radiology, 2008. 68(3): p. 362-368.
13. Bammer, R., Basic principles of diffusion-weighted imaging. European journal of radiology, 2003. 45(3): p. 169-184.
14. Gatehouse, P.D. and G.M. Bydder, Magnetic Resonance Imaging of Short T2 Components in Tissue. Clinical Radiology, 2003. 58(1): p. 1-19.
15. Haase, A., Snapshot flash mri. applications to t1, t2, and chemical-shift imaging. Magnetic Resonance in Medicine, 1990. 13(1): p. 77-89.
16. Le Bihan, D., et al., MR imaging of intravoxel incoherent motions: application to diffusion and perfusion in neurologic disorders. Radiology, 1986. 161(2): p. 401-407.
17. Nishimura, D.G., Time-of-flight MR angiography. Magnetic Resonance in Medicine, 1990. 14(2): p. 194-201.
18. R. Muthupillai, D.J.L., P.J. Rossman, J.F. Greenleaf, A. Manduca, R.L. Ehman, Magnetic resonance elastography by direct visualization of propagating acoustic strain waves. Science, 1995. 269: p. 1854-1857.
19. R. Muthupillai, R.L.E., Magnetic resonance elastography. Nature Medicine, 1996. 2: p. 601-603.
20. TO, W., Guidance for Industry and FDA Staff - Establishing Safety and Compatibility of Passive Implants in the Magnetic Resonance (MR) Environment, in http://www.fda.gov/cdrh/osel/guidance/1685.html August 21, 2008.
21. Karp, G.M.a.J.S., Positron emission tomography Phys. Med. Biol., 2006. 51: p. R117-R137.
22. Phelps, M.E., et al., Application of Annihilation Coincidence Detection to Transaxial Reconstruction Tomography. J Nucl Med, 1975. 16(3): p. 210-224.
23. Manning, K., et al., Clinical Practice Guidelines for the Utilization of Positron Emission Tomography/Computed Tomography Imaging in Selected Oncologic Applications: Suggestions from a Provider Group. Molecular Imaging and Biology, 2007. 9(6): p. 324-332.
24. Aarsvold, M.N.W.a.J.N., Emission Tomography: The Fundamentals of PET and SPECT. 2004: Elsevier Academic Press.
25. Nestle, U. and et al., Biological imaging in radiation therapy: role of positron emission tomography. Physics in Medicine and Biology, 2009. 54(1): p. R1.
26. Knoll, G.F., Single-photon emission computed tomography. Proceedings of the IEEE, 1983. 71(3): p. 320-329.
27. Mullan BP, O.C.M., Hung JC., Single photon emission computed tomography. Neuroimaging Clin N Am. , 1995. 5(4): p. 647-673.
28. Szabo, T.L., Diagnostic Ultrasound Imaging: Inside Out. 2004: Elservier Academic Press.
29. Hebden, S.R.A.a.J.C., Optical imaging in medicine: II. Modelling and reconstruction. Physics in Medicine and Biology, 1997. 42(5): p. 841-853.
30. Boss, A., et al., Hybrid PET/MRI of Intracranial Masses: Initial Experiences and Comparison to PET/CT. J Nucl Med, 2010: p. jnumed.110.074773.

31. *Boss, A., et al., Feasibility of simultaneous PET/MR imaging in the head and upper neck area. European Radiology, 2011: p. 1-8.*
32. *Bouchelouche, K., et al., PET/CT Imaging and Radioimmunotherapy of Prostate Cancer. Seminars in Nuclear Medicine, 2011. 41(1): p. 29-44.*
33. *Brunetti, J., et al., Technical Aspects of Positron Emission Tomography/Computed Tomography Fusion Planning. Seminars in Nuclear Medicine, 2008. 38(2): p. 129-136.*
34. *Cherry, S.R., Multimodality Imaging: Beyond PET/CT and SPECT/CT. Seminars in Nuclear Medicine, 2009. 39(5): p. 348-353.*
35. *Scharf, S., SPECT/CT Imaging in General Orthopedic Practice. Seminars in Nuclear Medicine, 2009. 39(5): p. 293-307.*
36. *Seo, Y., C. Mari, and B.H. Hasegawa, Technological Development and Advances in Single-Photon Emission Computed Tomography/Computed Tomography. Seminars in Nuclear Medicine, 2008. 38(3): p. 177-198.*
37. *Townsend, D.W., Positron Emission Tomography/Computed Tomography. Seminars in Nuclear Medicine, 2008. 38(3): p. 152-166.*
38. *Townsend, D.W., Multimodality imaging of structure and function. Physics in Medicine and Biology, 2008. 53(4): p. R1.*
39. *Pichler, B.J., et al., Positron Emission Tomography/Magnetic Resonance Imaging: The Next Generation of Multimodality Imaging? Seminars in Nuclear Medicine, 2008. 38(3): p. 199-208.*
40. *Mah, D. and C.C. Chen, Image Guidance in Radiation Oncology Treatment Planning: The Role of Imaging Technologies on the Planning Process. Seminars in Nuclear Medicine, 2008. 38(2): p. 114-118.*
41. *Patton, J.A., D.W. Townsend, and B.F. Hutton, Hybrid Imaging Technology: From Dreams and Vision to Clinical Devices. Seminars in Nuclear Medicine, 2009. 39(4): p. 247-263.*
42. *G Wang, H.Y., YB Ye, A scheme for multisource interior tomography. Medical Physics, 2009. 36(8): p. 3575-3581.*
43. *Jiansheng Yang, H.Y., Ming Jiang and Ge Wang, High-order total variation minimization for interior tomography. Inverse Problems, 2010. 3(26): p. 1-29.*
44. *Wang, H.Y.a.G., Compressed sensing based interior tomography. Physics in Medicine and Biology, 2009. 54(9): p. 2791-2805.*
45. *Yu HY, Y.J., Jiang M, Wang G, Interior SPECT- Exact and stable ROI reconstruction from uniformly attenuated local projections. Communications in Numerical Methods in Engineering, 2008: p. 18 pages.*
46. *J. Zhang, H.Y., C. Corum, M. Garwood, G. Wang Exact and Stable Interior ROI Reconstruction for Radial MRI. Proc. SPIE, 2009. Vol.7258: p. 8.*
47. *C H Moon, H.W.P., M H Cho and S Y Lee, Design of convex-surface gradient coils for a vertical-field open MRI system. Meas. Sci. Technol., 2000. 11(8): p. N89-N94.*
48. *While, P.T., L.K. Forbes, and S. Crozier, 3D gradient coil design for open MRI systems. Journal of Magnetic Resonance, 2010. 207(1): p. 124-133.*
49. *Laskaris, E.T., Open MRI magnet wth superconductive shielding. 1995, General Electric Company: United States.*
50. *Ye, Y., et al., A general local reconstruction approach based on a truncated Hilbert transform. International Journal of Biomedical Imaging, 2007. 2007: p. Article ID: 63634, 8 pages.*
51. *Yu, H. and G. Wang, Compressed sensing based Interior tomography. Phys Med Biol, 2009. 54(9): p. 2791-2805.*
52. *Yu, H., et al., Supplemental analysis on compressed sensing based interior tomography. Phys Med Biol, 2009. 54(18): p. N425-N432.*
53. *Yu, H., Y. Ye, and G. Wang, Local Reconstruction Using the Truncated Hilbert Transform via Singular Value Decomposition. Journal of X-Ray Science and Technology, 2008. 16(4): p. 243-251.*
54. *Boles, C.D., et al., A multimode digital detector readout for solid-state medical imaging detectors. Solid-State Circuits, IEEE Journal of, 1998. 33(5): p. 733-742.*
55. *Tumer, T.O.C., V.B. Clajus, M. Hayakawa S Volkovskii, A. , Multi-Channel Front-End Readout IC for Position Sensitive Solid-State Detectors. Nuclear Science Symposium Conference Record (NSS/MIC), 2006 IEEE Oct. 29 2006-Nov. 1 2006. 1: p. 384 - 388*
56. *Annie M. Tang, D.F.K., Edmund Y. Lam, Michael Brodsky , Ferenc A. Jolesz, Edward S. Yang Multi-modal Imaging: Simultaneous MRI and Ultrasound Imaging for Carotid Arteries Visualization, in 29th Annual International Conference of the IEEE Engineering in Medicine and Biology Society. 2007 Lyon.*
57. *Courdurier, M., et al., Solving the interior problem of computed tomography using a priori knowledge. Inverse Problems, 2008. 24: p. Article ID 065001 , 27 pages.*
58. *Kudo, H., et al., Tiny a priori knowledge solves the interior problem in computed tomography. Phys. Med. Biol., 2008. 53(9): p. 2207-2231.*
59. *Han, W., H. Yu, and G. Wang, A total variation minimization theorem for compressed sensing based tomography. International Journal of Biomedical Imaging, 2009. 2009: p. Articel ID:125871, 3 pages.*
60. *Yang, J.S., et al., High-order total variation minimization for interior tomography. Inverse Problems, 2010. 26(3): p. 29.*

61. Yu, H. and G. Wang, *Finite Detector Based Projection Model for Super Resolution CT*, in Fully3D 2011. 2011.
62. Donoho, D.L., *Compressed sensing.* IEEE Transactions on Information Theory, 2006. 52(4): p. 1289-1306.
63. Candes, E.J., J. Romberg, and T. Tao, *Robust uncertainty principles: Exact signal reconstruction from highly incomplete frequency information.* IEEE Transactions on Information Theory, 2006. 52(2): p. 489-509.
64. Chen, G.H., J. Tang, and S. Leng, *Prior image constrained compressed sensing (PICCS): A method to accurately reconstruct dynamic CT images from highly undersampled projection data sets.* Medical Physics, 2008. 35(2): p. 660-663.
65. Yu, H. and G. Wang, *A soft-threshold filtering approach for reconstruction from a limited number of projections.* Phys Med Biol, 2010. 55(13): p. 3905-3916.
66. Wang, G. and M. Jiang, *Ordered-Subset Simultaneous Algebraic Reconstruction Techniques (OS-SART).* Journal of X-ray Science and Technology, 2004. 12(3): p. 169-177.
67. Daubechies, I., M. Fornasier, and I. Loris, *Accelerated Projected Gradient Method for Linear Inverse Problems with Sparsity Constraints.* Journal Of Fourier Analysis And Applications, 2008. 14(5-6): p. 764-792.
68. Beck, A. and M. Teboulle, *A Fast Iterative Shrinkage-Thresholding Algorithm for Linear Inverse Problems.* Siam Journal on Imaging Sciences, 2009. 2(1): p. 183-202.
69. Fessler, J.A., *Model-Based Image Reconstruction for MRI.* IEEE Signal Processing Magazine. 27(4): p. 81-89.
70. Goldstein, T. and S. Osher, *The split Bregman algorithm for $l_1$ regularized problems.* SIAM J. Imaging Sci., 2009. 2: p. 323-343.
71. Robson, M.D., et al., *Magnetic Resonance: An Introduction to Ultrashort TE (UTE) Imaging.* Journal of Computer Assisted Tomography, 2003. 27(6): p. 825-846.
72. Idiyatullin, D., et al., *Fast and quiet MRI using a swept radiofrequency.* Journal of Magnetic Resonance, 2006. 181(2): p. 342-349.
73. Yu, H.Y., et al., *Interior SPECT-exact and stable ROI reconstruction from uniformly attenuated local projections.* Communications in Numerical Methods in Engineering, 2009. 25(6): p. 693-710.
74. Yang, J., et al., *High Order Total Variation Minimization for Interior SPECT.* Inverse Problems, 2011: p. Pending Revision.
75. Rullgard, H., *An explicit inversion formula for the exponential Radon transform using data from 180 ~.* Ark. Mat., 2004. 42: p. 353-362.
76. Cesareo, R.a.S.M., *A New Tomographic Device Based on the Detection of Fluorescent X-Rays.* Nuclear Instruments & Methods in Physics Research Section a-Accelerators Spectrometers Detectors and Associated Equipment, 1989. 277(2-3): p. 669-672.
77. Minchin, R.F.a.D.J.M., *Minireview Nanoparticles for Molecular Imaging-An Overview.* Endocrinology, 2010. 151(2): p. 474-481.
78. T. Yuasa, M.A., T. Takeda, M. Kazama, A. Hoshino, Y. Watanabe, K. Hyodo, F. A. Dilmanian, T. Akatsuka, and Y. Itai, *Reconstruction method for fluorescent x-ray computed tomography by least-squares method using singular value decomposition.* IEEE Trans. Nucl. SCI., 1997. 44: p. 54-62.
79. Teboulle, A.B.a.M., *A Fast Iterative Shrinkage-Thresholding Algorithm for Linear Inverse Problems.* Siam Journal on Imaging Sciences, 2009. 2: p. 183-202.
80. Bruce H. Hasegawa, E.L.G., Susan M. Reilly, Soo-Chin Liew and Christopher E. Cann, *Description of a simultaneous emission-transmission CT system.* 1990(Proc. SPIE 1231).
81. Beyer T, T.D., Brun T, et al., *A combined PET/CT scanner for clinical oncology.* J Nucl Med., 2000. 41: p. 1369-1379.
82. Jiang, M., et al., *Image reconstruction for bioluminescence tomography from partial measurement.* Opt Express, 2007. 15(18): p. 11095-116.
83. Cong, W., et al., *Practical reconstruction method for bioluminescence tomography.* Opt Express, 2005. 13(18): p. 6756-71.
84. Wang, G., Y. Li, and M. Jiang, *Uniqueness theorems in bioluminescence tomography.* Med Phys, 2004. 31(8): p. 2289-99.
85. Ntziachristos, V., et al., *Looking and listening to light: the evolution of whole-body photonic imaging.* Nat Biotechnol, 2005. 23(3): p. 313-20.
86. John J. Neumann, D.W.G.a.I.J.O. *Comparison of piezoresistive and capacitive ultrasonic transducers.* 2004.
87. Hassibi, A., A. Babakhani, and A. Hajimiri, *A Spectral-Scanning Nuclear Magnetic Resonance Imaging (MRI) Transceiver.* Solid-State Circuits, IEEE Journal of, 2009. 44(6): p. 1805-1813.
88. Rofougaran, A., et al., *A single-chip 900-MHz spread-spectrum wireless transceiver in 1-μm CMOS. I. Architecture and transmitter design.* Solid-State Circuits, IEEE Journal of, 1998. 33(4): p. 515-534.
89. Wang, G., H. Yu, and Y. Ye, *A scheme for multi-source interior tomography.* Med Phys, 2009. 36(8): p. 3575-3581.
90. Kok, P., et al., *Articulated planar reformation for change visualization in small animal imaging.* IEEE Trans Vis Comput Graph, 2010. 16(6): p. 1396-404.

91. *Khmelinskii, A., et al., Articulated Whole-Body Atlases for Small Animal Image Analysis: Construction and Applications. Mol Imaging Biol, 2010.*
92. *Baiker, M., et al., Atlas-based whole-body segmentation of mice from low-contrast Micro-CT data. Med Image Anal, 2010. 14(6): p. 723-37.*
93. *Osher, S., et al., An Iterative Regularization Method for Total Variation-Based Image Restoration. Multiscale Modeling & Simulation, 2005. 4(2): p. 460-489.*
94. *Osher, S. and J.A. Sethian, Fronts propagating with curvature-dependent speed: Algorithms based on Hamilton-Jacobi formulations. Journal of Computational Physics, 1988. 79(1): p. 12-49.*
95. *Osher, S.J.a.F., R. , Level Set Methods and Dynamic Implicit Surfaces, in Applied Mathematical Science. 2002, Springer.*
96. *Pfeiffer, F., et al., X-ray dark-field and phase-contrast imaging using a grating interferometer. Journal of Applied Physics, 2009. 105(10): p. 102006-102006-4.*
97. *Jin, Y., et al., Multifunctional nanoparticles as coupled contrast agents. Nat Commun, 2010. 1: p. 41.*
98. *Wang, G., Tomogranphy – Interior Tomographysiome (a provisional patent applicaed filed; VTIP 11-103, Application Number 61471245). 2011.*
99. *Wang, G., H. Gao, and J. Zhang, MRI Based on an Inhomogeneous Magnetic Field (Patent Disclosure filed with Virginia Tech, and a provisional patent application filed as well). 2011.*

Dr. Ge Wang received his MS and PhD degrees in electrical and computer engineering from State University of New York, Buffalo in 1991 and 1992, respectively. He was on the faculty with Mallinckrodt Institute of Radiology, Washington University, St. Louis, MO, from 1992 to 1996, and with University of Iowa, Iowa City, IA, from 1997 to 2006. He is currently the Samuel Reynolds Pritchard Professor and the Director of Biomedical Imaging Division, Wake Forest University – Virginia Tech (WFU–VT) School of Biomedical Engineering and Sciences, Virginia Tech, Blacksburg, VA. His interests include x-ray computed tomography (CT), optical molecular tomography, and inverse problems. He is the author or coauthor of about 300 journal articles and numerous conference papers, including the first paper on spiral/helical cone-beam CT (a main mode of modern CT scanners with ~100 million scans annually in USA), the first paper on bioluminescence tomography (BLT), and the first paper on interior tomography. He is the founding Editor-in-Chief of the *International Journal of Biomedical Imaging*. Dr. Wang is Associate Editor for IEEE Trans. on Medical Imaging and IEEE Trans. on Biomedical Engineering. He is a Fellow of the Institute of Electrical and Electronics Engineers (IEEE), Society of Photographic Instrumentation Engineers (SPIE), Optical Society of America (OSA), and American Institute for Medical and Biological Engineering (AIMBE) (http://www.imaging.sbes.vt.edu).

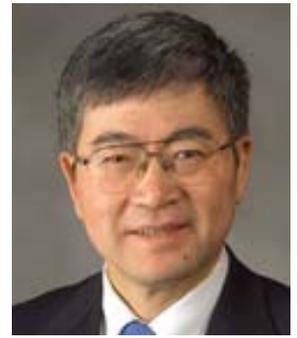

Dr. Jie Zhang received his PhD degrees in Biomedical Engineering from Tianjin University, China in 1999 and in Biophysical Sciences and Medical Physics from University of Minnesota, Minneapolis in 2004, respectively. He was a research fellow at the Mayo Clinic, Rochester from 2004 to 2006. He is currently an assistant professor, and an American Board of Radiology (ABR) certified medical physicist with Department of Radiology, University of Minnesota. Dr. Zhang has extensive experience on various imaging modalities and their clinical applications. He has also conducted large amount of research in clinical medical imaging and physics, brain function, physiological measurement and modeling, and impedance imaging. His work has been published in various peer-viewed journals such as Stroke, Radiology, Academic Radiology, Physiological Measurement, etc. His research interests include medical imaging (MRI, CT, Ultrasound, EIT, and etc.), biomedical modeling, physiological measurement and image and signal processing.

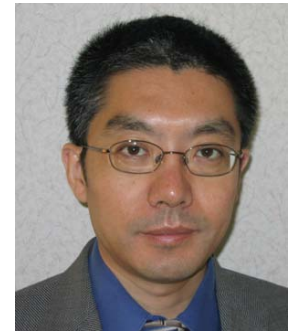

Dr. Hao Gao received his B.Sc. in astronomy and astrophysics from University of Science and Technology of China, Hefei, China in 2004, his M.S. in medical physics from University of California, Irvine in 2005, and his Ph.D. in applied mathematics from University of California, Irvine in 2010. He is currently an assistant adjunct professor in Department of Mathematics of University of California, Los Angeles. His research interests include inverse problems, optimization, numerical analysis, scientific computing, optical tomography, x-ray computed tomography, magnetic resonance imaging, compressive and rapid reconstruction techniques for medical imaging.

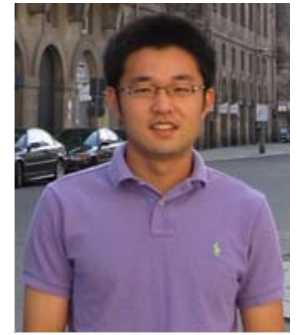

Dr. Victor Weir received his PhD in Biophysical Sciences and Medical Physics from the University of Minnesota, Minneapolis in 2008. He was assistant professor/Assistant Clinical Specialist at the University of Minnesota until February 2009. He is currently a Medical Physicist in the Baylor Health Care System in Dallas, Texas. His interests include the physics of biological systems - from molecular structure to species, x-ray Computed Tomography imaging and dosimetry, radionuclide imaging, and the modeling of electronic structure and optical properties of materials. He has coauthored articles in both peer review journals and conference proceedings. Dr. Weir has also served as a reviewer for the IEEE Engineering in Medicine and Biology Conferences (EMBC). His memberships in professional organizations include the American Association of Physicists in Medicine (AAPM), and the Society of Nuclear Medicine (SNM).

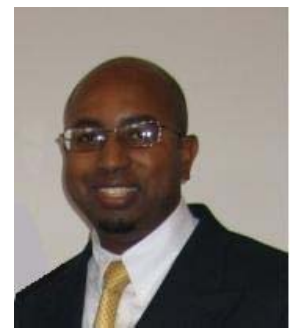

Dr. Hengyong Yu received Bachelor's degrees in information science & technology (1998), computational mathematics (1998), and PhD in information & telecommunication engineering (2003) from Xi'an Jiaotong University, Xi'an, China. He was on the faculty with the College of Telecommunication Engineering, Hangzhou Dianzi University, Hangzhou, China, from 2003 to 2004. He was a research faculty with the University of Iowa, Iowa City, IA, from 2004 to 2006, and with the Virginia Tech, Blacksburg, VA, from 2006 to 2010. Currently, He is an Assistant Professor, the Director of CT Lab, Departments of Radiology and Biomedical Engineering, Wake Forest University Health Sciences, Winston-Salem, NC. His interests include computed tomography and medical image processing. He has authored or coauthored >80 peer reviewed journal papers. He serves as the Editorial Board members of Signal Processing, CT Theory and Applications, International Journal of Biomedical Engineering and Consumer Health Informatics and Open Medical Imaging Journal, and Guest Editor of the International Journal of Biomedical Imaging for the special issue entitled "Development of Computed Tomography Algorithms". He is a senior member of the IEEE and a member of the Biomedical Engineering Society.

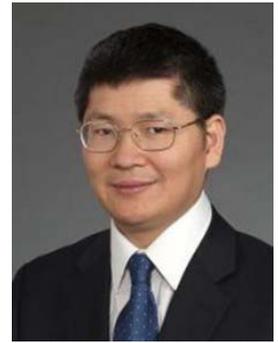

Dr. Wenxiang Cong is currently a research assistant professor with Biomedical Imaging Division, School of Biomedical Engineering and Sciences, Virginia Tech, Blacksburg. He has published over 50 peer-reviewed journal articles. His research interests include optical molecular tomography, x-ray phase-contrast and dark-field imaging.

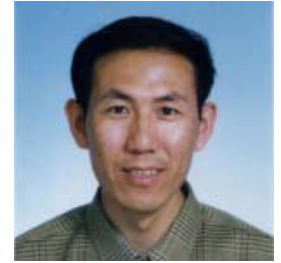

Dr. Xiaochen Xu was born in Hangzhou, China, in 1976. He received the B.A.Sc. in Electrical Engineering from Nanjing University, Nanjing, China, in 1999, and the M.Sc. degree in Acoustics from the same university in 2002. He received his Ph.D. degree in Biomedical Engineering from the University of Southern California, Los Angeles, in 2007. His research interests included the design of beam formers for high-frequency ultrasonic arrays and high-frequency Doppler ultrasound for medical applications. Dr. Xu joined the Medical Business Unit at Texas Instruments in June, 2007. Currently he works as an applications manager and systems engineer at TI and is developing highly integrated circuit for ultrasound applications.

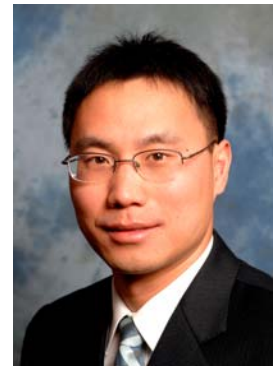

Dr. Haiou Shen is currently a research scientist with Biomedical Imaging Division, School of Biomedical Engineering and Sciences, Virginia Tech, Blacksburg. He has published over 30 peer-reviewed journal articles. His research interests include optical molecular tomography and Monte-Carlo simulation.

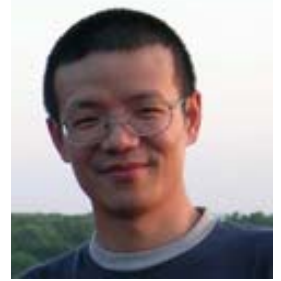

James Bennett received his BS in Biomedical Engineering from the University of Iowa in 2005. He was an IT Consultant at Accenture, Ltd. (Chicago, IL) for four years before starting his PhD in 2010 at Virginia Tech–Wake Forest University School of Biomedical Engineering & Sciences, Virginia Tech, Blacksburg, VA. He is also the IT Manager for the Biomedical Imaging Division. His interests include diagnostic imaging, cardiac/vascular imaging, x-ray computed tomography (CT), multi-spectral x-ray imaging and IT infrastructure.

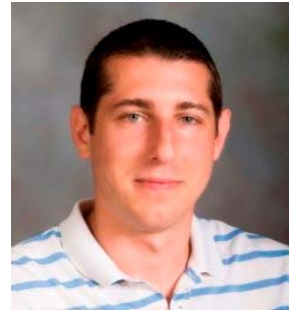

Dr. Yue Wang received his B.S. and M.S. degrees in electrical and computer engineering from Shanghai Jiao Tong University in 1984 and 1987 respectively. He received his Ph.D. degree in electrical engineering from University of Maryland Graduate School in 1995. In 1996, he was a postdoctoral fellow at Georgetown University School of Medicine. From 1996 to 2003, he was an Assistant and later Associate Professor of Electrical Engineering at The Catholic University of America. In 2003, he joined Virginia Tech and is currently the Grant A. Dove Professor of Electrical and Computer Engineering and director of Computational Bioinformatics and Bio-imaging Laboratory. His research interests focus on pattern recognition, machine learning, statistical signal/image processing, with applications to computational bioinformatics and biomedical imaging for human disease research. He is the author or coauthor of about 80 journal articles, with papers appeared in IEEE Transactions, Nature Reviews, Nature Medicine, and Bioinformatics. He is a Fellow of the American Institute for Medical and Biological Engineering (AIMBE) and ISI Highly Cited Researcher by Thomson Scientific (yuewang@vt.edu;http://www.cbil.ece.vt.edu).

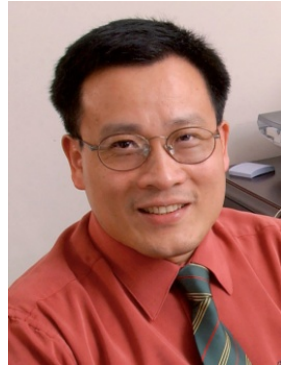

Dr. Michael Vannier is currently a professor with Department of Radiology, University of Chicago. He is an expert in diagnostic radiology, specializing in gastrointestinal imaging. Dr. Vannier's research interests include medical imaging modalities, visualization, and image processing. His research has been funded by the National Institutes of Health. He holds six U.S. patents, including one for gastrointestinal tract unraveling and two for computer-based upper extremity evaluation. He has authored numerous book chapters and more than 200 articles in peer-reviewed journals. Dr. Vannier serves on the editorial boards of several medical journals including the *International Journal of Computer Aided Radiology and Surgery* and the *American Journal of Orthopaedics*. Dr. Vannier is a member of the National Aeronautics and Space Administration (NASA)/U.S. Space Foundation Hall of Fame.

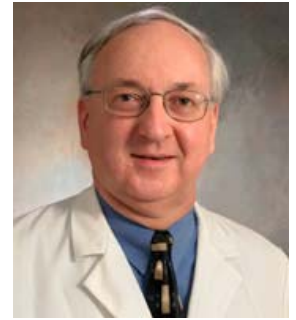

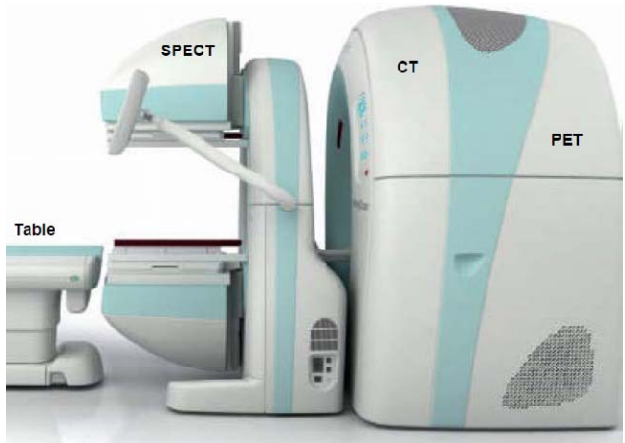 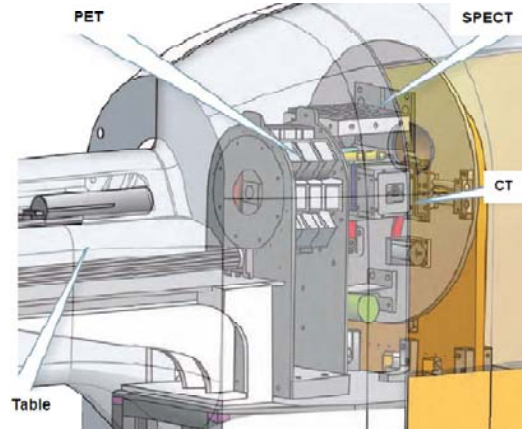

(a)  (b)

**Figure I.C.1.** State of the art tri-modality fusion systems. (a) The AnyScan system for clinical PET-SPECT-CT, and (b) the Albira system for preclinical PET-SPECT-CT ((a) and (b) from http://www.mediso.de/anyscan-sc.html and http://www.cmi-marketing.com/7modalities respectively, with the legends added by the authors).

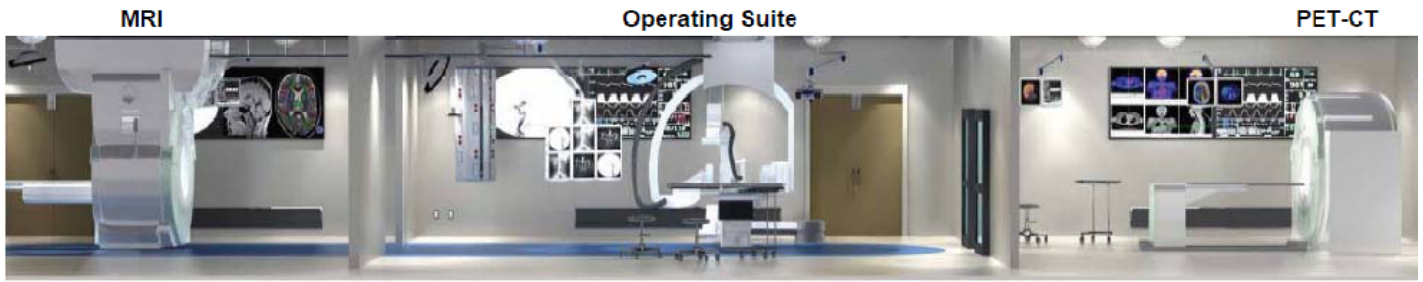

**Figure I.D.1.** Multimodality image guided operating suite. "*On May 4, 2011, Brigham and Women's Hospital (BWH) will unveil the Advanced Multimodality Image Guided Operating (AMIGO) Suite, the first multimodality suite in the world to give surgeons and interventional specialists immediate access to a full array of imaging modalities for use during procedures*" (from the AMIGO News Letter Issue 1, March 2011, with the legends added by the authors of this paper).

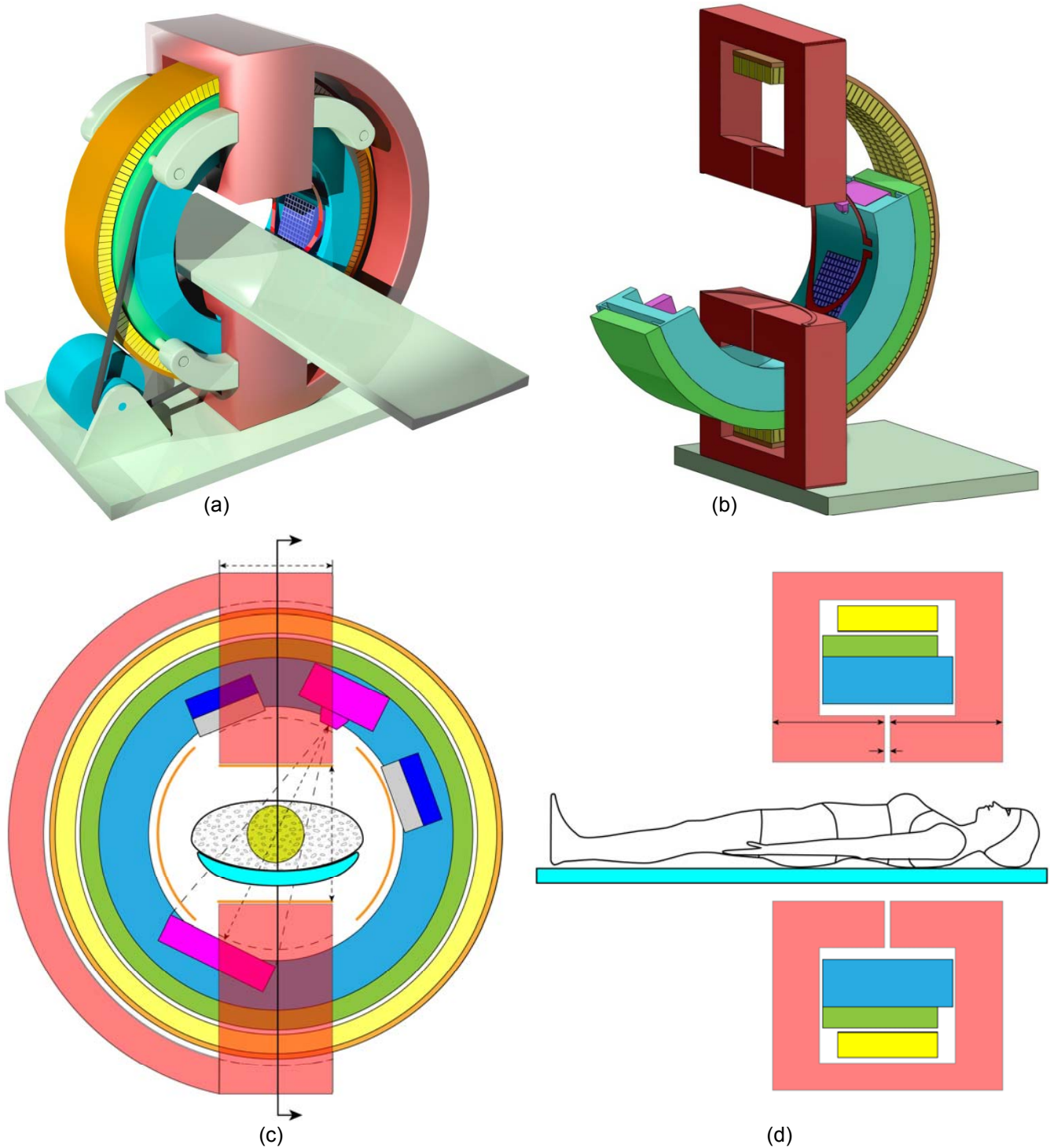

**Figure II.A.1.** Exemplary multi-tomography/omni-tomography system architecture. (a) A 3D rendering of the multi-tomography top-level design, (b) a partial rendering, (c) an in-plane view, and (d) a through-plane view. There are two static rings and one rotating ring for multi-tomography. While the inner ring is a permanent magnet and the outer ring contains PET crystals, the middle ring supports a CT tube, a CT detector and a pair of SPECT detectors. The middle ring is on a slip ring (like a large ball bearing) as the interface for both CT and SPECT. The rotating ring, the slip-ring, and the PET ring all go through the magnetic poles.

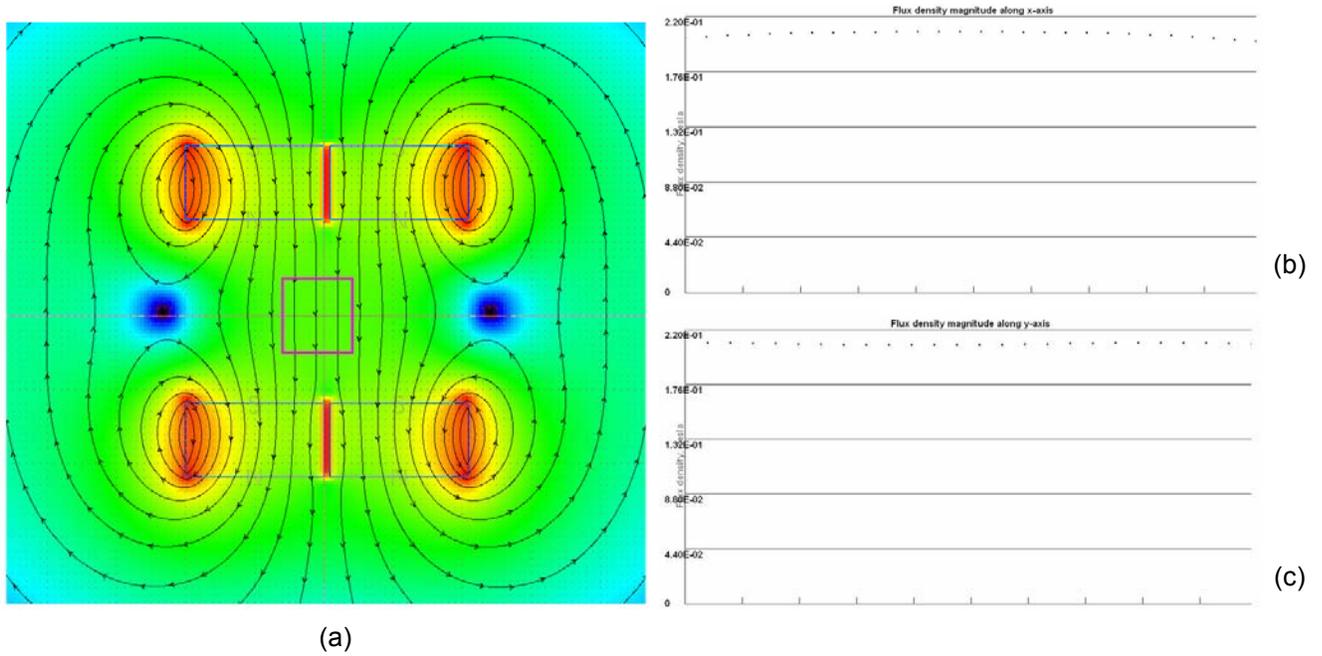

**Figure II.B.1.** Generation of a local magnetic field. (a) The magnetic flux from a simple arrangement of four permanent blocks. The square area (20 x 20 cm$^2$) represents a region of interest (ROI) where the magnetic flux ranges from 0.208 to 0.211 tesla; (b) and (c) the magnetic flux plots along the x- and y-axes respectively. Each magnetic block is of 40 x 40 x 20 cm$^3$, with a gap of 2 cm between two parts of each magnetic pole.

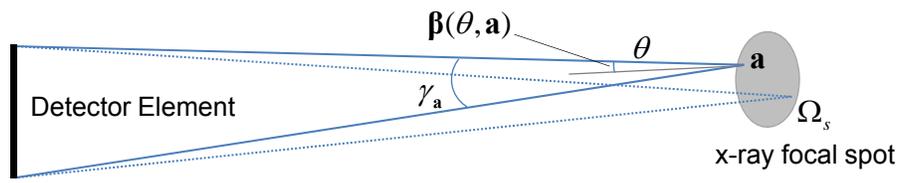

**Figure III.A.1.** Projection model with a finite detector element and a finite focal spot in fan-beam geometry.

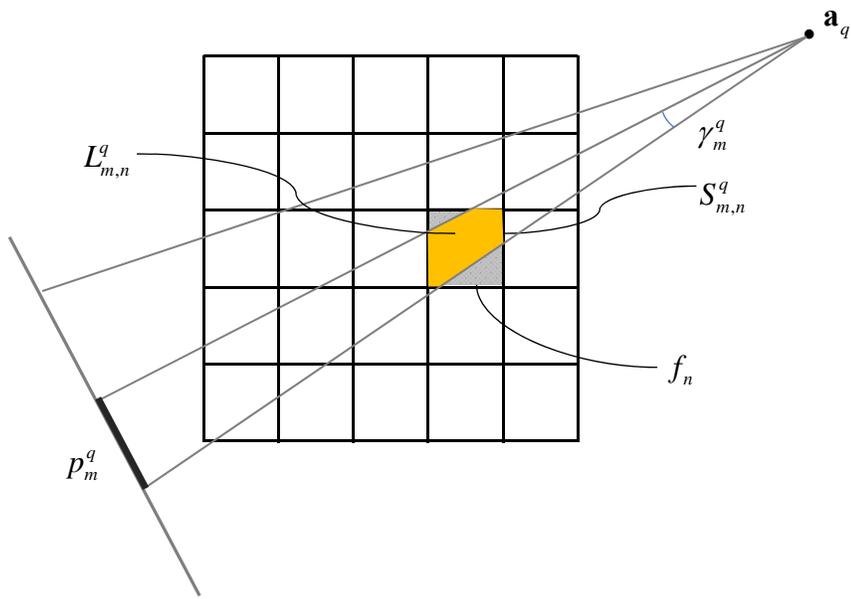

**Figure III.A.2.** Discrete projection model assuming a discrete image in fan-beam geometry.

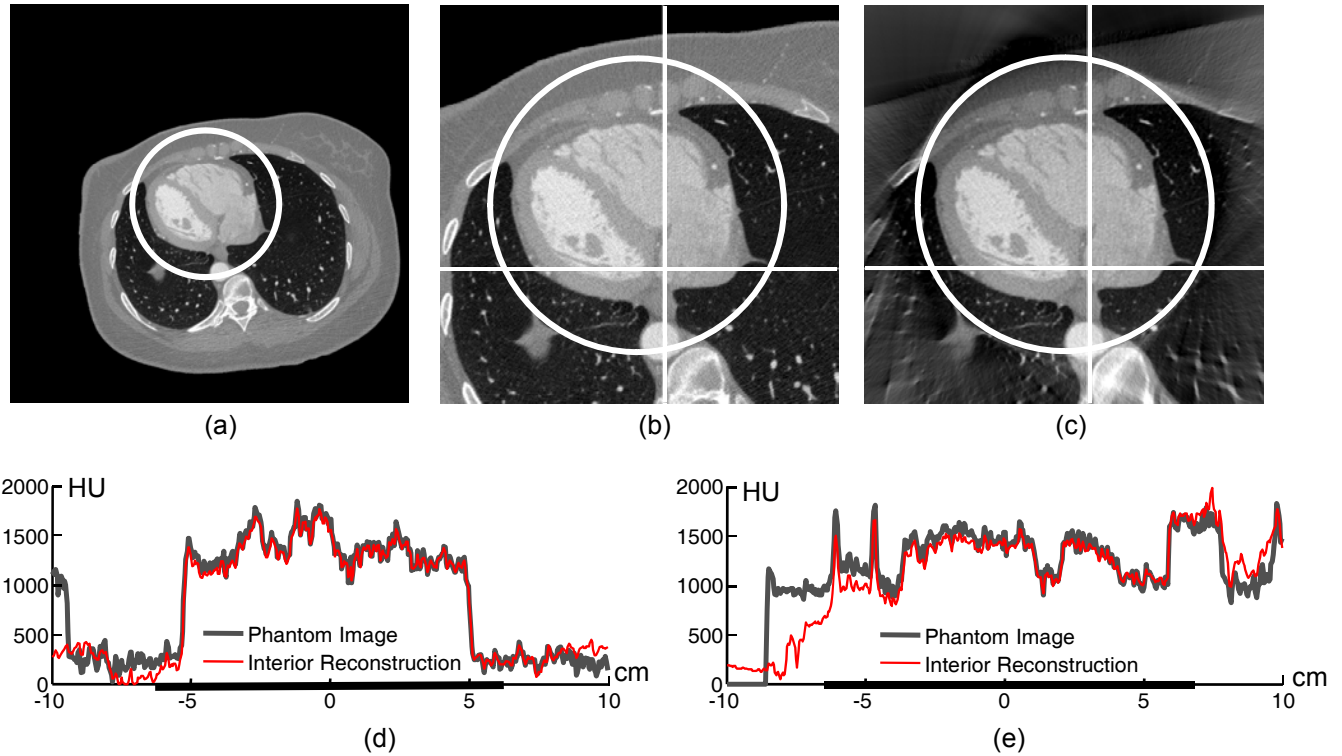

**Figure III.A.3.** Interior CT reconstruction. (a) A realistic CT image of 1,000 by 1,000 pixels covering an area of 40X40 cm$^2$ with a cardiac region of interest (ROI); (b) a magnified ROI of 20X20 cm$^2$; (c) the corresponding interior reconstruction after 15 iterations; (d) and (e) the profiles along the horizontal and vertical white lines in (b) and (c) respectively, where the thick lines on the horizontal axis indicate the ROI. The display window for (a)-(c) is [0,2000].

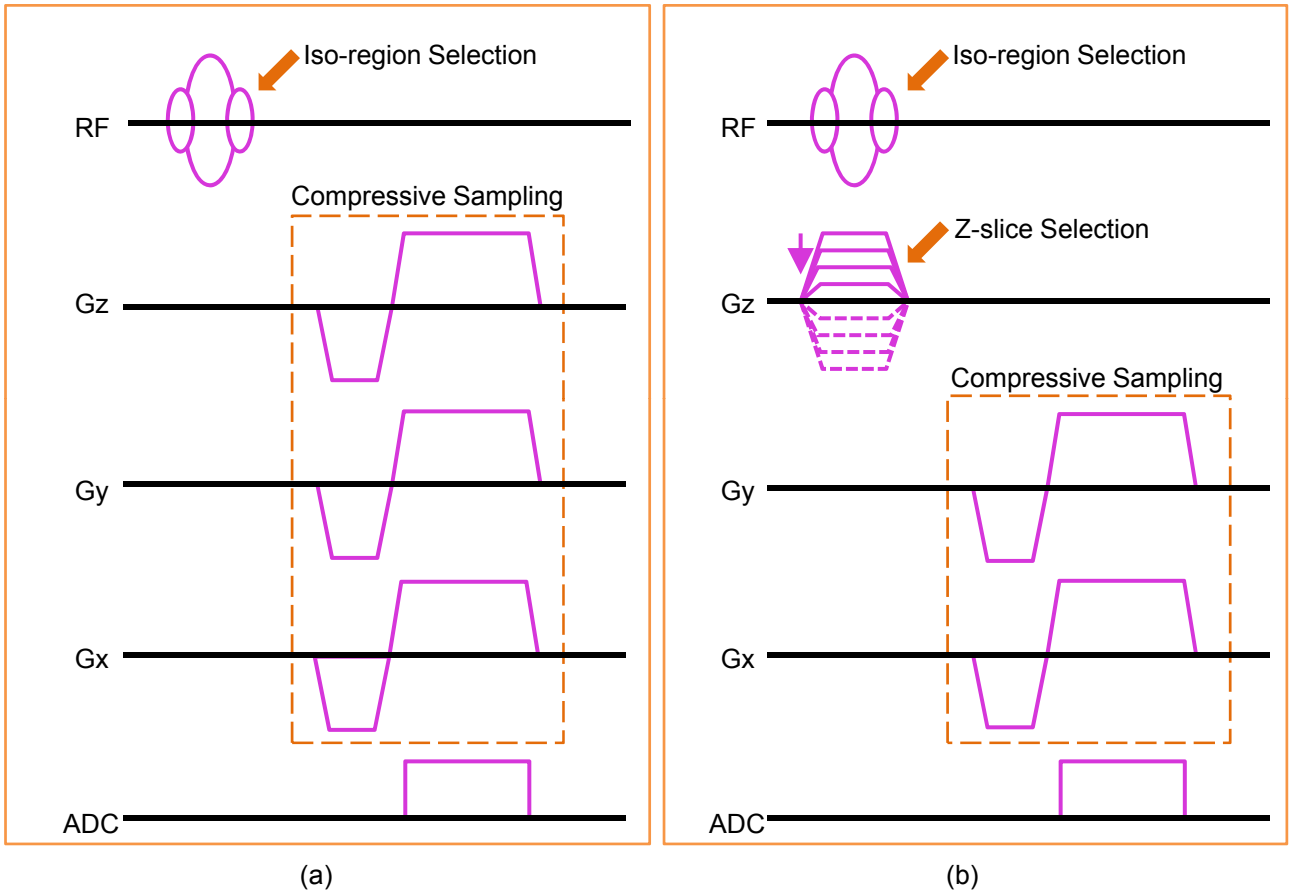

**Figure III.B.1.** Pulse sequences for interior MRI based on an inhomogeneous magnetic field. (a) An illustrative sequence for 3D imaging, and (b) a sequence for slice-based imaging.

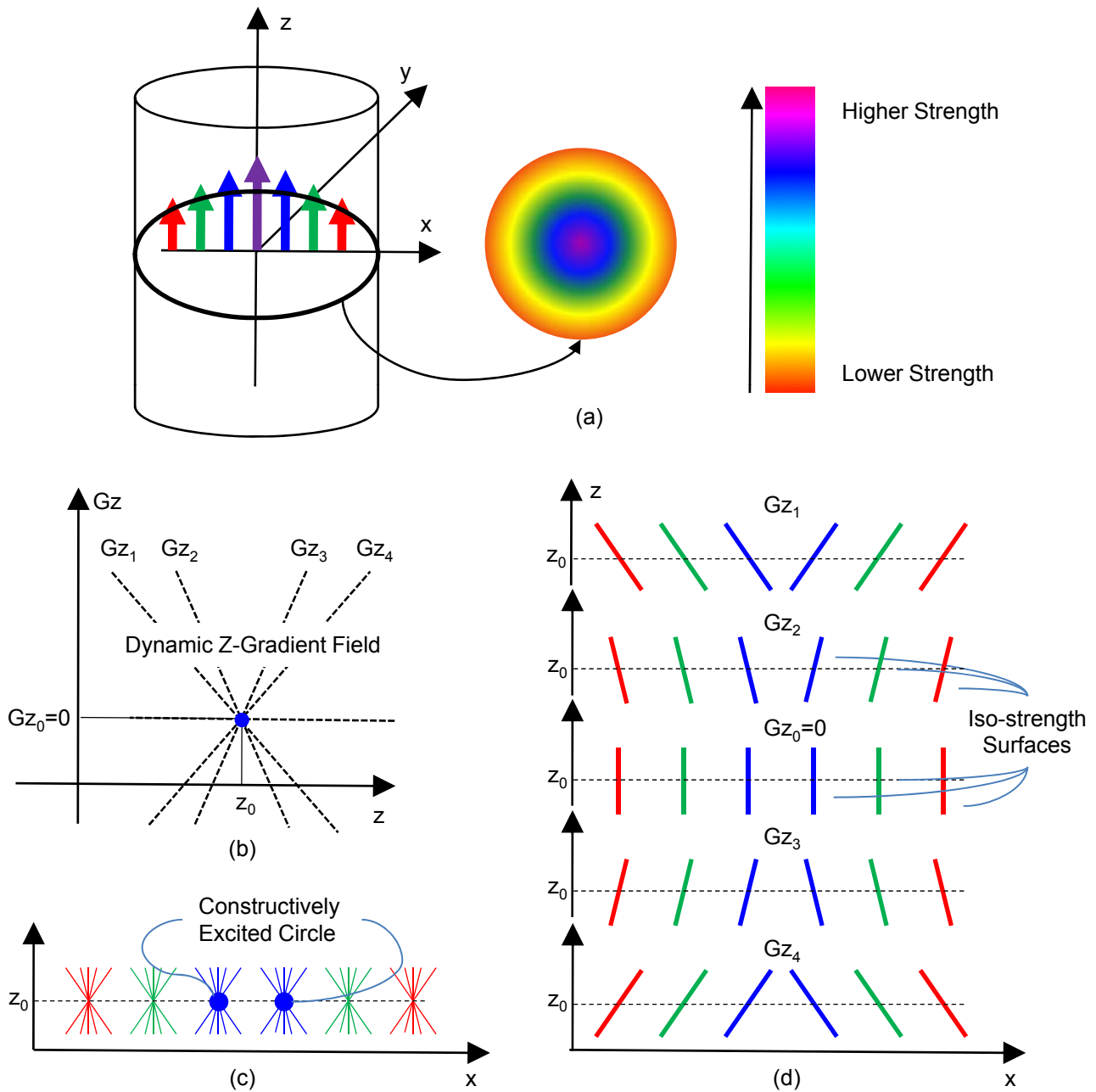

**Figure III.B.2.** Slice selection using a dynamic z-gradient field. (a) A circularly symmetric magnetic field whose magnitude is the same along the z direction but linearly decreasing from the z axis; (b) a time-varying z-gradient field is applied upon the inhomogeneous magnetic background field; (c) only a specified transverse circle in a patient is constructively excited; and (d) the iso-strength surfaces in the composite magnetic field is a function of the z-gradient field, which helps explain the constructive excitation mechanism shown in (c). Although only five z-gradient fields are shown, we may use many more in practice or change the z-gradient continuously.

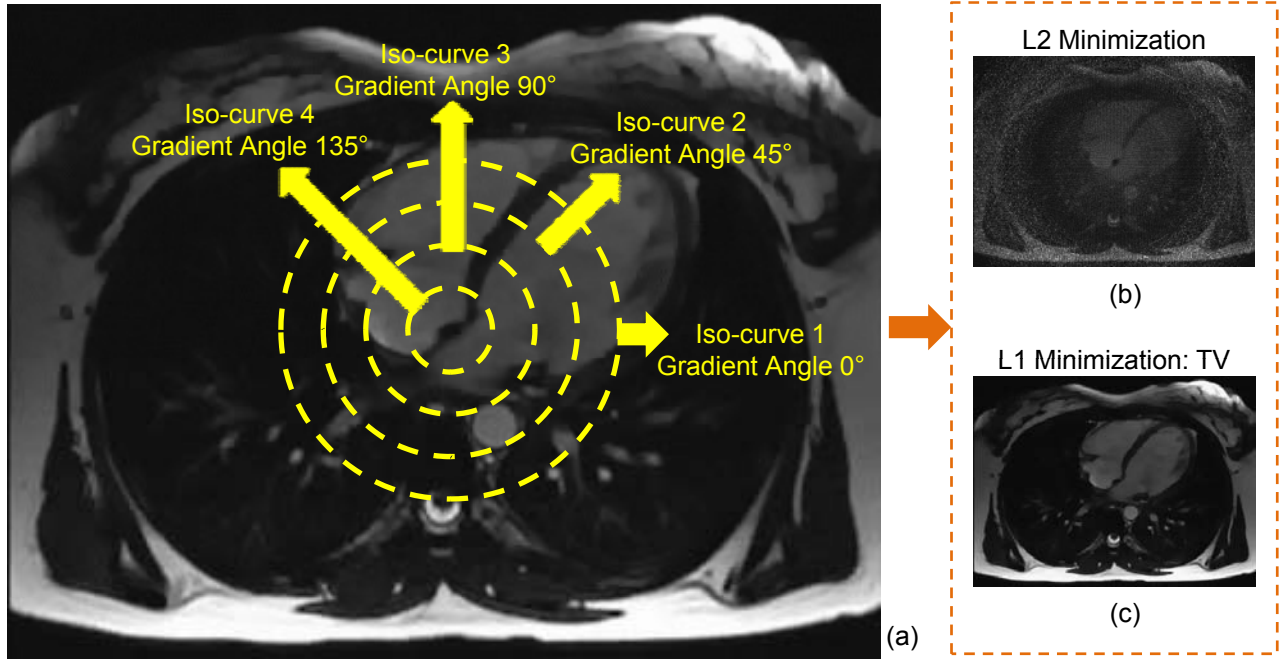

**Figure III.B.3.** Global MRI reconstruction. (a) A compressive sensing scheme, (b) a L2 reconstruction, and (c) a L1 reconstruction in terms of TV minimization from a global MRI dataset.

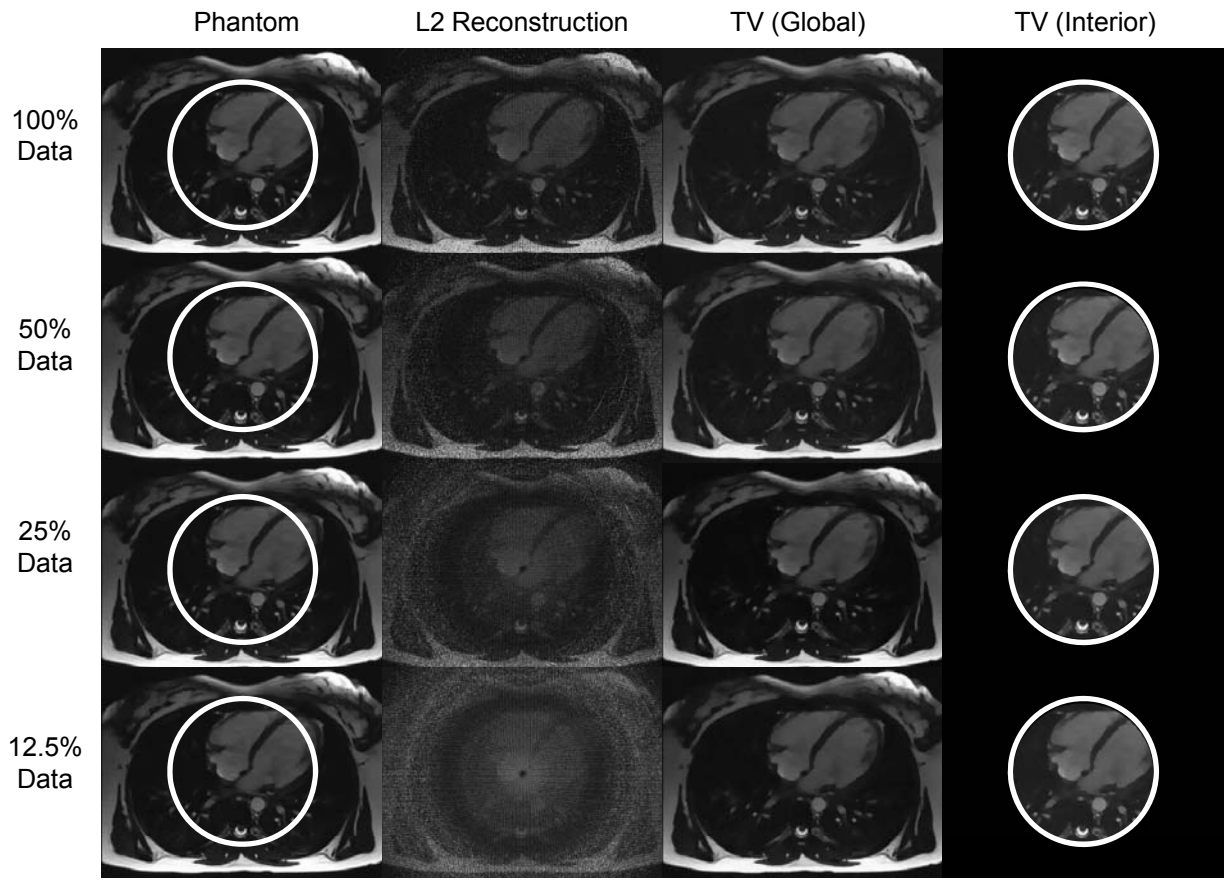

**Figure III.B.4.** Interior MRI reconstruction using the L2- and L1(TV)- minimization from under-sampled and ROI-based datasets. The sampling is based on the scheme illustrated in **Figure III.B.3.**

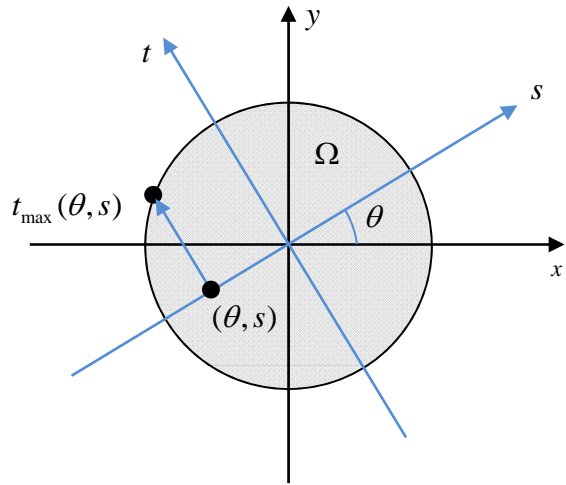

**Figure III.C.1.** Interior SPECT imaging in parallel-beam geometry.

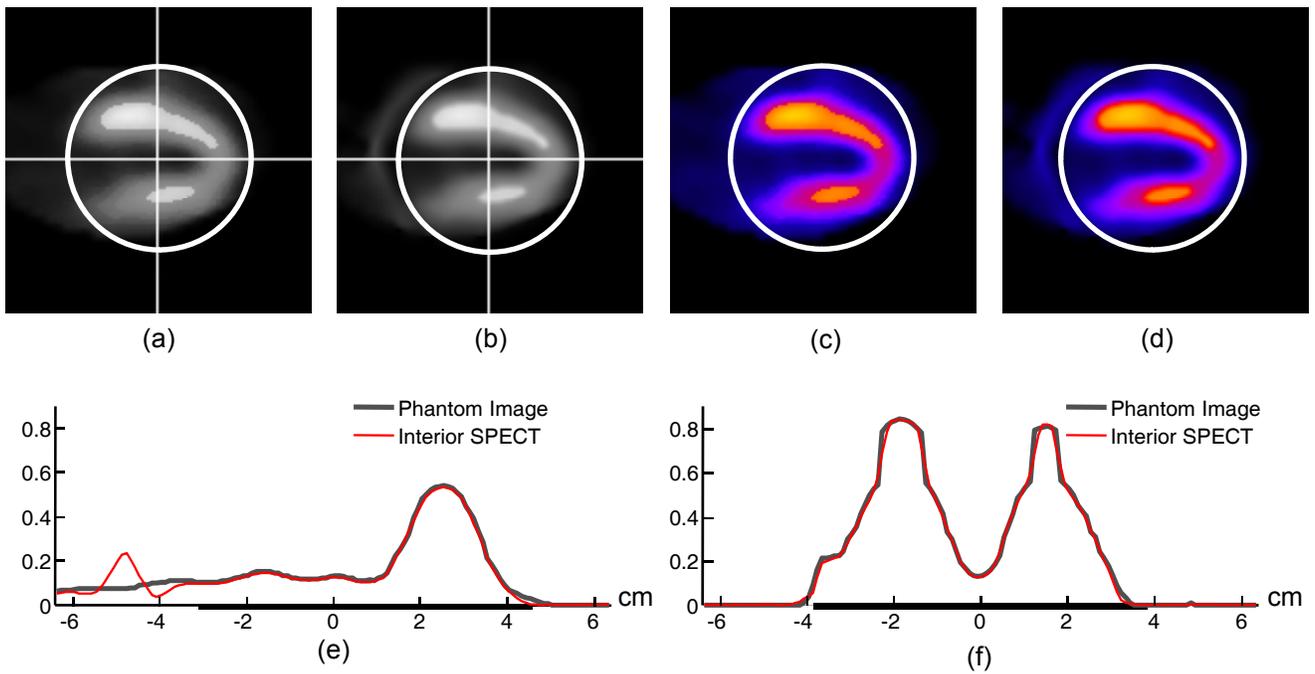

**Figure III.C.2.** Interior SPECT reconstruction. (a) An original SPECT ROI image of 128 by 128 pixels covering an area of 12.8X12.8 cm²; (b) an interior reconstruction using the HOT minimization algorithm with the attenuation coefficient $\mu_0$=0.15 after 40 iterations; (c) and (d) the pseudo-color counterparts of (a) and (b) respectively; and the representative profiles along the (e) horizontal and (f) vertical white lines respectively. The horizontal axis in (e) and (f) represents the 1D coordinate, and the vertical axis denotes the functional value. The thick lines on the horizontal axis indicate the ROI. The display window for (a) and (b) is [0 1.0].

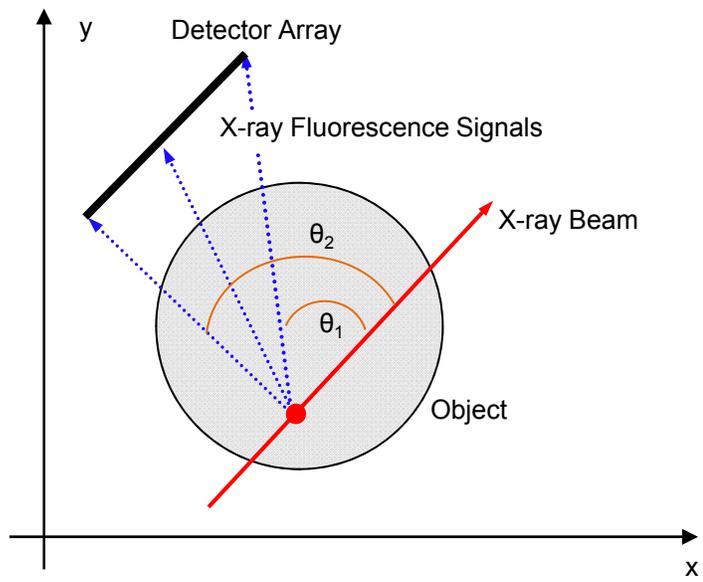

**Figure III.D.1.** Imaging geometry for x-ray fluorescence CT.

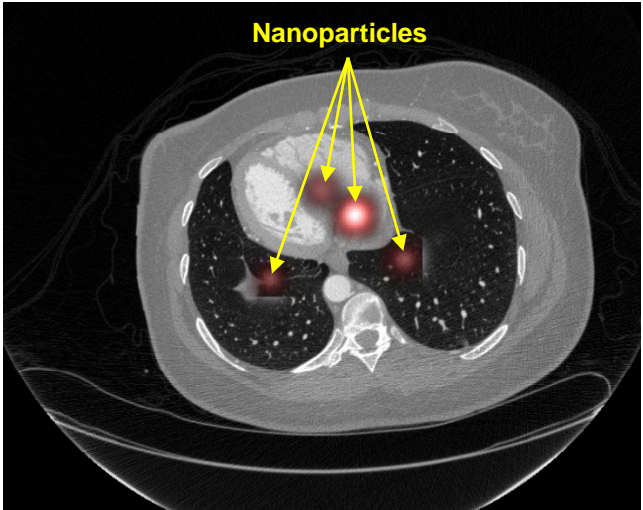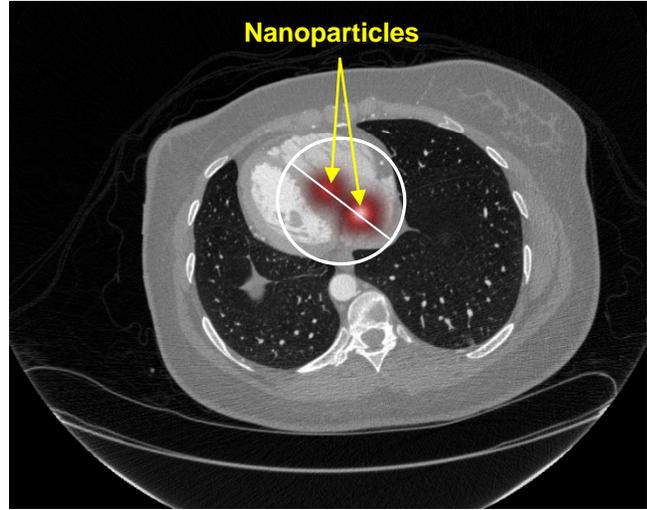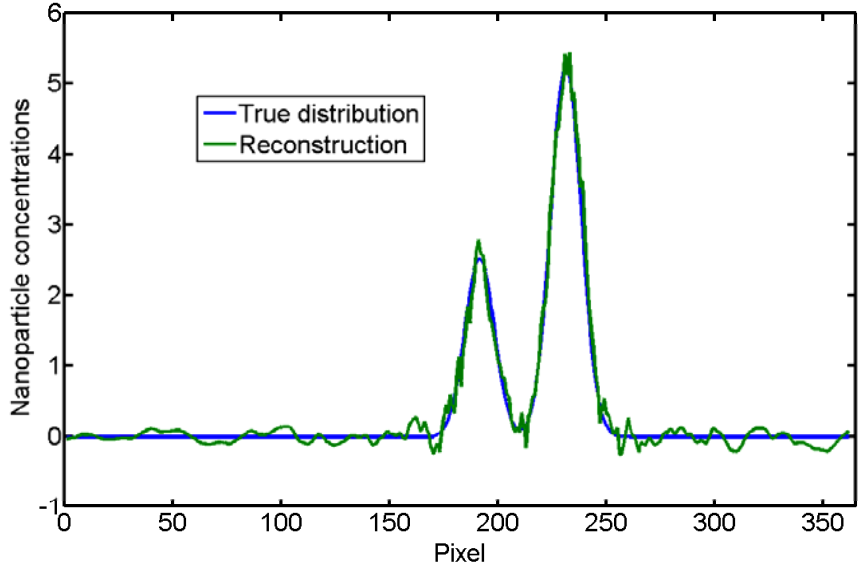

**Figure III.D.2.** Interior XFCT reconstruction. (a) The true clusters of golden nanoparticles accumulated in four circular regions overlapped on a CT image, (b) the reconstructed nanoparticle clusters in the ROI, and (c) the profiles through the oblique line shown in (b).